\documentclass[superscriptaddress,aps,preprintnumbers,amsmath,amssymb,prd,nofootinbib,twocolumn]{revtex4-2}

\usepackage{graphicx}
\usepackage{epstopdf}
\usepackage{dcolumn}
\usepackage{bm}
\usepackage{hyperref}
\usepackage{color}
\usepackage{amsmath}
\usepackage{braket}
\usepackage{ulem}

\def\df{\mathrm{d}}
\def\gst{{\rm g}_*}

\def\Re{\mathrm{Re}}

\renewcommand{\emph}[1]{\textit{#1}}

\begin{document}

\title{
Stochastic effects on observation of ultralight bosonic dark matter
}

\author{Hiromasa Nakatsuka}
\affiliation{Institute for Cosmic Ray Research, University of Tokyo, Kashiwa 277-8582, Japan}
\author{Soichiro Morisaki}
\affiliation{Department of Physics, University of Wisconsin-Milwaukee, Milwaukee, WI 53201, USA}
\author{Tomohiro Fujita}
\affiliation{Waseda Institute for Advanced Study, Waseda University, Shinjuku, Tokyo 169-8050, Japan}
\affiliation{Research Center for the Early Universe (RESCEU), The University of Tokyo, Bunkyo, Tokyo 113-0033, Japan}
\author{Jun'ya Kume}
\affiliation{Department of Physics, University of Tokyo, Bunkyo, Tokyo 113-0033, Japan}
\affiliation{Research Center for the Early Universe (RESCEU), The University of Tokyo, Bunkyo, Tokyo 113-0033, Japan}
\author{Yuta Michimura}
\affiliation{Department of Physics, University of Tokyo, Bunkyo, Tokyo 113-0033, Japan}
\affiliation{PRESTO, Japan Science and Technology Agency (JST), Kawaguchi 332-0012, Japan}
\author{Koji Nagano}
\affiliation{Institute of Space and Astronautical Science, Japan Aerospace Exploration Agency, Sagamihara City 252-5210, Japan}
\author{Ippei Obata}
\affiliation{Max-Planck-Institut f{\"u}r Astrophysik, Karl-Schwarzschild-Straße. 1, 85748 Garching, Germany}

\begin{abstract}

Ultralight bosonic particles are fascinating candidates of dark matter (DM).
It behaves as classical waves in our Galaxy due to its large number density.
There have been various methods proposed to search for the wave-like DM, such as methods utilizing interferometric gravitational-wave detectors.
Understanding the characteristics of DM signals is crucial to extract the properties of DM from data.
While the DM signal is nearly monochromatic with the angular frequency of its mass, the amplitude and phase are gradually changing due to the velocity dispersion of DMs in our Galaxy halo.
The stochastic amplitude and phase should be properly taken into account to accurately constrain the coupling constant of DM from data.
Previous works formulated a method to obtain the upper bound on the coupling constant incorporating the stochastic effects.
One of these works compared the upper bound with and without the stochastic effect in a measurement time that is much shorter than the variation time scale of the amplitude and phase.
In this paper, we extend their formulation to arbitrary measurement time and evaluate the stochastic effects.
Moreover, we investigate the velocity-dependent signal for dark photon DM including an uncertainly of the velocity.
We demonstrate that our method accurately estimates the upper bound on the coupling constant with numerical simulations.
We also estimate the expected upper bound of the coupling constant of axion DM and dark photon DM from future experiments in a semi-analytic way.
The stochasticity especially affects constraints on a small mass region.
Our formulation offers a generic treatment of the ultralight bosonic DM signal with the stochastic effect.

\end{abstract}

\maketitle
\tableofcontents

\section{Introduction}
\label{sec_intro}

These days, various experiments are planned and performed to investigate the broad candidates of dark matter (DM) including ultralight bosonic DMs~\cite{Arvanitaki:2014faa,Graham:2015ifn,Geraci:2016fva,Budker:2013hfa,Stadnik:2014tta,Arvanitaki:2015iga,Arvanitaki:2016fyj,Vermeulen:2021epa} such as axionlike particles and dark photons.
Axionlike particles~\cite{Peccei:1977hh,Svrcek:2006yi} have a unique coupling to the photon, which rotates the polarization direction of the photons.
The polarization rotation provides a new detection scheme through the laser technique.
Hereafter, axionlike particles are called axion for short.
The dark photon is a massive vector field characterised by its $U_D(1)$-charge.
The baryon charge ($D=B$) or baryon minus lepton charge ($D=B-L$) induces the dark electric force on the experimental equipment like mirrors in interferometers, which can be detected by the gravitational wave interferometers without any additional modification~\cite{Pierce:2018xmy,Guo:2019ker,Morisaki:2020gui,Michimura:2020vxn,Miller:2020vsl}.
Such DM experiments by interferometers are highly sensitive to the axion and dark photon signals.
These experiments can improve the current constraints obtained by
other types of experiments, e.g. 
CAST~\cite{CAST:2004gzq,CAST:2017uph} and 
the astrophysical observations (SN1987A~\cite{Payez2015}, M87~\cite{Marsh2017}, and NGC 1275~\cite{Reynolds:2019uqt}) for the axion, 
and E\"ot-Wash experiment \cite{Schlamminger:2007ht, Wagner:2012ui}
and the MICROSCOPE experiment \cite{Touboul:2017grn, Berge:2017ovy, Fayet:2017pdp} for the dark photon
(see also Refs.~\cite{Pierce:2018xmy,Nagano:2019rbw,Morisaki:2020gui}).

Understanding the characteristics of the DM signal is crucial to distinguish the signal from various noises and extract the properties of DM from data~\cite{Foster:2017hbq,Derevianko:2016vpm,Centers:2019dyn,Miller:2020vsl}.
The DM signal is nearly monochromatic with the angular frequency of the DM mass, which enables us to extract the DM signal at the oscillation frequency. 
However, the amplitude and phase of the DM field are gradually changing due to the velocity dispersion of DMs in our Galaxy halo.
This time scale is called a ``coherent time $\tau$'' of ultralight DM.
The amplitude and phase appear to evolve in a stochastic manner over the coherent time.
It is known that this stochastic nature of ultralight DM suppresses the sensitivity of the observation in the following two points: 
(i) When the measurement time $T$ is longer than the coherent time $\tau$, the modulation of the phase broadens the DM signal in frequency space, which slows down the improvement of sensitivity with time~\cite{Budker:2013hfa,Miller:2020vsl}.
(ii)  When $T$ is shorter than $\tau$, we sample only one realization of the amplitude of DM, which could be a smaller value than the average one by chance.
Since the realized field value is random, the observed amplitude has uncertainty, which loosens the upper bound on the DM coupling constant about factor $\mathcal O(1)$~\cite{Centers:2019dyn}.
These previous works focused on either of two effects, and  the intermediate region $T\sim \tau$, where both effects are relevant, has not been investigated.
Although Ref.~\cite{Foster:2017hbq} considered both effects in their analysis, they did not compare the upper bound with and without the point (ii).
As interferometer experiments improve the sensitivity on a low-frequency range, the intermediate region becomes more important to search for the ultralight DM.
Moreover, the dark photon signal can have another uncertainty from its velocity dependence while its stochastic effect was not well studied.
Since these stochastic effects inevitably affect any experiments to detect the ultralight DM in the intermediate region, it is necessary to understand the characteristics of DM signals.

In this paper, we investigate the stochastic effect of ultralight bosonic DM.
The bosonic DM field consists of a superposition of DM particles that have slightly different velocities.
We evaluate the superposed waves in the frequency space to derive a probability distribution of the field value.
We also derive the probability distribution of the spatial derivative of a vector field for the first time, which characterizes a velocity-dependent signal of a dark photon DM.
Next, we formulate a frequentist's method to put an upper bound of a DM coupling constant.
We compare the upper bound with and without the stochastic effect of the amplitude.
We find that the stochastic effect becomes negligible as the measurement time sufficiently exceeds the coherent time.
We also confirm that our results are consistent with Ref.~\cite{Centers:2019dyn} for $T<\tau$.
Our method can be applied for generic experiments to estimate the upper bound on the ultralight bosonic DM.
As an application, we estimate the future upper bound by the  Advanced LIGO(aLIGO)-like experiment based on our method.

This paper is described in the following way.
In Sec.~\ref{sec_Stochastic_DM}, we revisit the stochastic nature of ultralight bosonic DM. 
We derive the probability distribution of the field value and its spatial derivative.
We formulate the DM signals of axion and dark photon in Sec.~\ref{sec_observation}.
In Sec.~\ref{sec_sensitivity}, we formulate a method to estimate the upper bound of the coupling constant taking into account the stochastic effects.
We derive the likelihood functions of the DM signals with experimental noise.
The probability distribution of DM signals is confirmed by numerical simulations.
We apply our formulation to estimate the constraints given by the future gravitational interferometer experiment for axion and dark photon DMs in Sec.~\ref{sec_experiment}.
We compare the upper bound with and without the stochastic effect of the DM amplitude to evaluate its contribution.
Sec.~\ref{sec_conclusion} is devoted to the conclusion of this paper.
Throughout this paper, we employ natural units in which $\hbar=c=1$, unless otherwise specified.

\section{Stochastic nature of ultralight bosonic DM}
\label{sec_Stochastic_DM}

Ultralight DM behaves like a classical wave, which is contributed by multiple waves with different frequencies and phases.   
Firstly, we review the formulation of ultralight DM and its stochastic time-evolution.
Next, we perform the Fourier transform of the field value to derive its power spectrum, which is the main feature in ultralight DM searches.

\subsection{Review on the properties of ultralight DM}
\label{sec_review_ultra_light_DM}

Cold DM consists about 25\% of our Universe, which has the local density $\rho_{\rm DM}\sim 0.4 {\rm GeV/cm}^3$ and a virial velocity $v_{\rm vir} \simeq 220~{\rm km/sec } \simeq 7\times 10^{-4}c$ around the solar system in our Galaxy~\cite{Bertone:2004pz,Evans:2018bqy}.
When the mass of DM is ultralight, $m\ll 1$~eV, its number density is extremely large as 
\begin{align}
    &n_\phi L_\phi^3 
    =
    \frac{\rho_{\rm DM}}{ m^4 v_{\rm vir}^3}
    \nonumber
    \\&=
    8\times 10^{51} 
    \frac{\rho_{\rm DM}}{0.4 ~{\rm GeV/cm}^3}
    \left(\frac{10^{-12}~{\rm eV}}{m}\right)^4
    \left(\frac{220~{\rm km/sec } }{v_{\rm vir}}\right)^3
    ,\label{UDM number density}
\end{align}
where $n_\phi \equiv \rho_{\rm DM}/m$ is the number density and $L_\phi \sim 1/(v_{\rm vir}m)$ is the de Broglie wavelength.
Thus, the ultralight DM should be a boson, not a fermion.
In this paper, we focus on a ultralight bosonic DM, $m\ll 1\mathrm{eV}$, such as axionlike particle $\phi=a$ and dark photon $\phi=A_x, A_y, A_z$.
Note that the temporal component of a dark photon, $A_0$, is negligibly small compared to spatial components, and therefore we ignore $A_0$ in our calculation~\cite{Michimura:2020vxn}.
We can treat ultralight bosonic DM as a classical wave since it has a large number density.

The DM velocity at the surface of the Earth has some components, e.g. the virial velocity of a DM halo of our Galaxy, the velocity of the Sun to the halo rest frame, the orbital motion of the Earth, and the rotation of the Earth.
We consider two dominant components, a virial velocity $\vec v_h$ and the velocity of the Sun $|\vec v_{\odot}| \simeq 232 $~km/sec.
We define the velocity of DM as $\vec v = \vec v_h- \vec v_{\odot}$,
where $\vec v_{\odot}$ is a constant and $\vec v_h$ is a random variable following the probability distribution of the standard halo model~\cite{Evans:2018bqy}:
\begin{align}
	f_{\rm SHM}(\vec v_h) ~\df^3 \vec v_h
	&= \frac{1}{(\pi v_{\rm vir}^2)^{3/2}}
	\exp\left[- \frac{(\vec v_h)^2}{ v_{\rm vir}^2}\right]
    ~\df^3 \vec v_h.
    \label{eq_SHM_vec}
\end{align}
The distribution of DM speed is given by 
\begin{align}
    & \overline f_\mathrm{SHM}(v) ~\df v
    \equiv
    \df v~ v^2\int \df^2\Omega_e
    f_{\rm SHM}(v\vec e + \vec v_\odot)
    \nonumber
	\\&= 
	\frac{v}{\sqrt{\pi} v_{\rm vir} v_{\odot} }
	\exp\left[- \frac{(v+v_{\odot})^2}{v_{\rm vir}^2}\right]
	\left(
	e^{4v v_{\odot} /v_{\rm vir}^2 }   -1
	\right)
	~\df v,
    \label{eq_SHM_abs}
\end{align}
where $\int \df^2\Omega_e$ represents the integration over a direction of a unit vector $\vec e$.
The  typical velocity of DM is given by
\begin{align}
    \bar v ^2
    &\equiv 
    \int \df v ~\overline f_\mathrm{SHM}(v)
    v^2 
    \nonumber
    \\&=
    \int \df^3 \vec v_h~
    f_{\rm SHM}(\vec v_h)  (\vec v_h- \vec v_{\odot})^2
    = v_{\odot} ^2 + {\frac{3}{2}}v_{\rm vir}^2.
    \label{def_barv}
\end{align}

The DM field is contributed by many partial waves with different velocities following the distribution $\bar{f}_{\mathrm{SHM}}(v)$.
The total field value $\Phi(t,\vec x)$ is given by the sum of $N_\phi$ partial waves as
\begin{align}
    \Phi(t,\vec x)
    &=
    \sigma_\phi N_\phi^{-1/2}
    \sum_{i=1}^{N_\phi}
    \cos\left(
    m(1+v_i^2/2)t 
    +m \vec v_i\cdot \vec x
    +\theta_i
    \right),
    \label{eq_phi_superposed}
\end{align}
where $\theta_i$ is a random phase of the $i$-th wave, $\vec v_i$ is its velocity, $v_i \equiv |\vec v_i|$, and $\sigma_\phi N_\phi^{-1/2}$ is a normalization constant.
Note that although the number of the superposed partial waves is huge in reality (see Eq.~\eqref{UDM number density}),
$\Phi(t,\vec x)$ well approximates the realistic field for a sufficiently large $N_\phi 
(\gg 1)$.
The normalization is determined by the energy density,
 $
 \rho_{\rm DM}/{\rm g}_*\simeq V^{-1}\int_V \df x^3
     \frac{1}{2}\left(
      \dot\Phi^2
     + m^2\Phi^2
     \right)
$,
 where $V$ is a volume we are considering and
 $\gst$ is the number of degrees of freedom,  $\gst=1$ for scalar field and $\gst=3$ for vector field.
The typical field value, $\sigma_\phi$, is written as
 \begin{align}
     \sigma_\phi
     \equiv
     \sqrt{
     \frac{2\rho_{\rm DM}}{g_*m^2 }
     },
 \end{align}
 where 
 $\sigma_a  \equiv \sqrt{2\rho_{\rm DM}/m^2  }$ for axion DM
 and 
  $\sigma_A  \equiv \sqrt{2\rho_{\rm DM}/(3m^2)  }$ for dark photon DM.
 
\begin{figure}[tb]
  \centering
  \includegraphics[width=0.33\columnwidth]{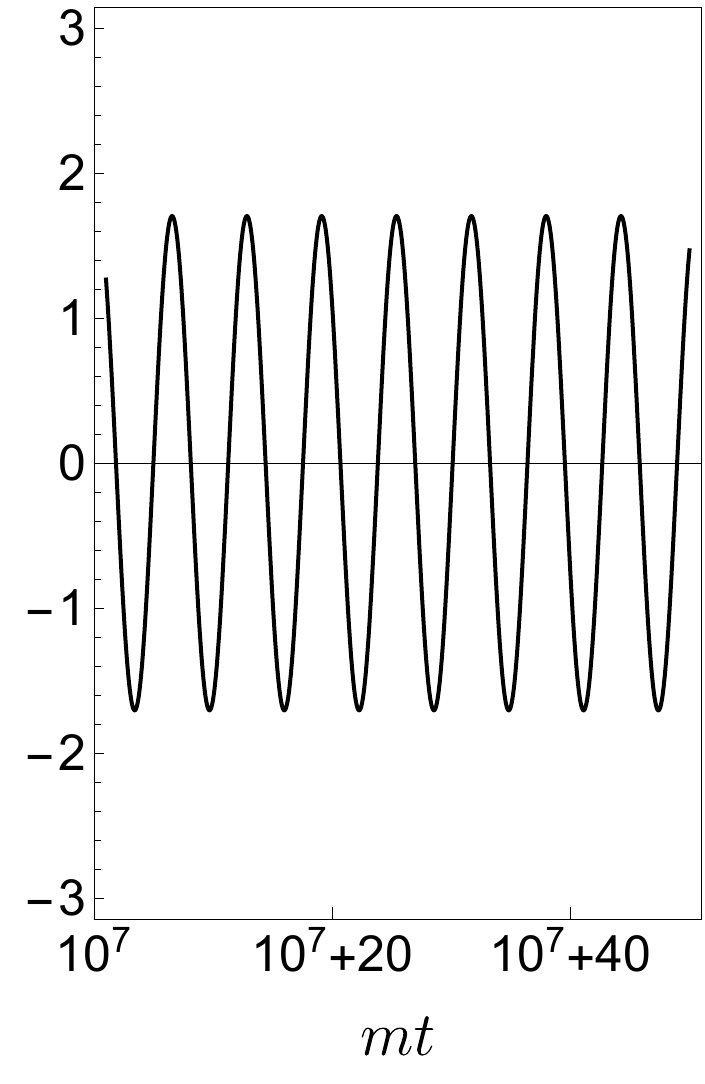}
  \includegraphics[width=0.65\columnwidth]{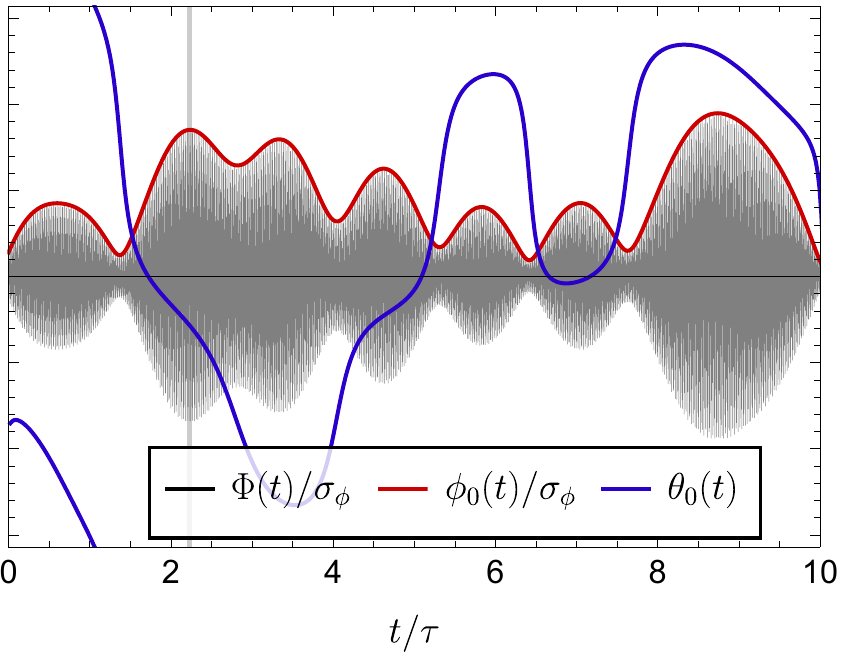}
  \caption{
  The time dependence of the amplitude and the phase of bosonic field.
  We numerically calculate a realization of superposed wave in Eq.~\eqref{eq_phi_superposed} following the method in Ref.~\cite{Foster:2017hbq}, 
  where we divide the angular frequency $\omega=m(1+v^2/2)$ by $10^4$ bins over $\omega/m\in[1, 1+5v_\mathrm{vir}^2]$.
  {The left panel presents the field oscillation for a shorter time scale 
  which corresponds to the vertical light gray band at $t\sim 2\tau$ in the right panel.}
  The time is normalized by the coherent time \eqref{eq_coherent_time}.
  The red line represents the normalized amplitude and the blue line represents the phase.
  }
  \label{fig_boson_mockdata}
\end{figure}

Each wave has a similar angular frequency $m(1+v_i^2/2) \sim m$ because typical velocity is small  $\bar{v}\sim 10^{-3}$. 
Consequently, the total amplitude and the oscillating phase of $\Phi$ are almost constant until the frequency difference between plane waves becomes significant, $\tau m \bar{v}^2/2 = \pi$. 
This time scale $\tau$ is called a coherent time, 
\begin{align}
    \tau \equiv 
    \frac{2\pi}{m\bar v^2}
    \simeq 4 \times 10^6 m^{-1}.
    \label{eq_coherent_time}
\end{align}
For a longer time scale than the coherent time $\tau$, the phase and amplitude of $\Phi$ slowly vary. 
In Fig.~\ref{fig_boson_mockdata}, the time evolution of the total field value $\Phi(t)$ is shown.
We numerically evaluate Eq.~\eqref{eq_phi_superposed} by setting $\vec x =0$ without loss of generality.
In the left panel, we show the time evolution of the field over several oscillatory periods at the gray vertical line in the right panel.
In the right panel, the field dynamics is shown for 
 $10\tau$.
The amplitude and phase noticeably change over the coherent time scale.
Here, we fit the field dynamics by $\Phi(t) = \phi_0\cos(m(1+\bar v^2/2)t+\theta_0)$ over several oscillatory periods around each time point to obtain the instant amplitude $\phi_0$ and the phase $\theta_0$ that are shown as red and blue lines.

\subsection{Formulation of stochastic field values}
\label{sec_formulation_stochastic_field}

Now we derive the power spectrum of the DM field value $\Phi$, which is often analyzed by an experiment searching for an ultralight scalar field to extract the periodic signal.
The power spectrum has two features~\cite{Foster:2017hbq}.
First, its spectral shape has some width corresponding to the velocity dispersion.
Second, its amplitude is a stochastic variable similar to the field value in a time coordinate.
Ref.~\cite{Foster:2017hbq,Centers:2019dyn}  estimated the probability distribution of the power spectrum of the field value.
In this paper, we extend their method to the spatial derivative of the field value, which  is relevant to the dark photon DM search as discussed in Sec.~\ref{sec_observation}.

Here, we derive the spectral shape and the probability distribution of the Fourier-transformed field value, time-derivative, and spatial derivative.
We choose $\vec x = 0$ without loss of generality since a shift of the coordinate just results in a constant phase of each partial wave.
Provided that we have the time-series data of the field value over the observation time $T$ spanning $t=[-T/2,T/2]$,
the frequency space is discretized as $f_n$ for the $n$-th bin with a resolution $\Delta f = 1/T$, that it, $f_n -f_{n-1}=1/T$.

At first, the Fourier-transformation of the field value~\eqref{eq_phi_superposed} is given by~\cite{Foster:2017hbq}
(See App.~\ref{app_fourier_trans_field_values} for derivation.)
\begin{align}
    \tilde\Phi(f_n)
    \nonumber
    &\equiv
    \int^{T/2}_{-T/2} \df t\, e^{-2\pi i f t}
    \Phi(t,\vec x)
    |_{\vec x= 0}
\\
   &\simeq
   \frac{T}{2}
   \sigma_\phi
   ~
    \sqrt{ \Delta_s(f_n)}
    ~
    \left[
   \frac{r_n}{{\sqrt{2}}}
   \exp(i\theta_n)
   \right]
   ,
    \label{eq_phi_fj_superposed}
\\
    \Delta_s(f_n)
\nonumber
    &=
    \int^{f_n+\Delta f/2}_{f_n-\Delta f/2} 
    \overline f_\mathrm{SHM}(v)
    \frac{\df v}{\df f} \df f
    \label{eq_def_Delta_N}
    \\
    &=
    \frac{1}{2}
    \bigg[
        \text{erf}\left(\frac{v-v_\odot}{v_\mathrm{vir}}\right)
        +\text{erf}\left(\frac{v+v_\odot}{v_\mathrm{vir}}\right)
\nonumber
    \\\quad&
    +\frac{
    v_\mathrm{vir}
    }{\sqrt{\pi } v_\odot}
     \left(
         e^{-\frac{(v +v_\odot)^2}{v_\mathrm{vir}^2}}
     -
         e^{-\frac{(v -v_\odot)^2}{v_\mathrm{vir}^2}}
     \right)
     \bigg]^{v(f_n+\Delta f/2)}_{v(f_n-\Delta f/2)}
    ,
\end{align}
where erf$(x)$ represents the error function, the phase $\theta_{n}$ is a stochastic variable following a uniform distribution over $[0,2\pi]$, and the amplitude $r_{n}$ is following the standard Rayleigh distribution,
\begin{align}
     P_{\rm R}(r)\df r = 
     r 
     \exp
     \left(
     -\frac{r^2}{2}
     \right)
     ~\df r
     .
      \label{eq_Rayleigh}
\end{align}
$\tilde\Phi(f_n)$ consists of three parts.
First, the coefficient $(T/2)\sigma_\phi$ represents the typical amplitude of the field value in Fourier space.
Second, $r_n$ and $\theta_n$ in $[\ldots]$ represent the stochastic fluctuations, which follow the standard Rayleigh distribution and the uniform distribution, respectively.
Third, $\Delta_s(f_n)$ characterizes the deterministic part of the spectral shape, which represents the  fraction of DM waves that belong to the $n$-th frequency bin.
Consequently, $ \Delta_s(f_n)$ is normalized so that 
\begin{align}
    \Delta_{s,\mathrm{tot}}
    \equiv 
    \sum_n \Delta_s(f_n)
    = 
    \int^{\infty}_{f_\mathrm{DM}} 
    \overline f_\mathrm{SHM}(v)
    \frac{\df v}{\df f} \df f
    =1
    ,
    \label{eq_delta_N_tot}
\end{align}
where $f_\mathrm{DM} \equiv m/(2\pi)$ is a frequency of the field oscillation, below which $\Delta_s(f_n) =0$.

Next, we consider the time derivative of the DM field value since some experiments search for the time modulation of the field value.
The Fourier transformation of field value, $\dot\Phi(t)\equiv (\mathrm{d} \Phi(t)/\df t)$, is given by
\begin{align}
    \int \df t e^{-2\pi i f_n t}
    \dot\Phi(t)
   &=
   2\pi f_n i
   \frac{T}{2}
   \sigma_\phi
    \sqrt{ \Delta_s(f_n)}
    ~
    \frac{r_n}{{\sqrt{2}}}
   \exp(i\theta_n)
   \nonumber
    \\&\simeq
     i m \tilde \Phi(f_n)
     \label{eq_dotPhi_formula}
    .
\end{align}
In the last equality, we use the fact that the DM signal appears only around the frequency of DM mass, $f_n\sim m/(2\pi)$.
Thus, the time derivative of the field value has a similar spectral distribution with $\tilde \Phi(f_n)$.

The spatial derivative of field value is more complicated.
The spatial derivative of field value on a direction $j~(=x,y,z)$ is given by,
(See App.~\ref{app_fourier_trans_field_values} for derivation.)
\begin{align}
    &\nabla_j \tilde \Phi
    (f_n,\vec x)
    \bigg|_{\vec x=0}
   =
   \frac{T}{2}
   \sigma_\phi
   m \bar v
   ~
    \sqrt{ \Delta_{j}(f_n)  }
    ~
    \left[
    \frac{r_n}{{\sqrt{2}}}
    \exp\left(
    i\theta_{n}
    \right)
    \right]
    ,
    \label{eq_def_Del_Phi}
    \\
     &\Delta_{j}(f_n)
    =
    \int^{f_n+\Delta f/2}_{f_n-\Delta f/2} 
     v^2
     \frac{\df v}{\df f} \df f
    \int \df^2\Omega_e
    \nonumber
    \\&\qquad\qquad
    \times
    f_{\rm SHM}(\vec v(f,\vec e) + \vec v_\odot)
    \frac{[v_j(f_n,\vec e_l)]^2 }{\bar v^2}
    ,
    \label{eq_def_Delta_v}
\end{align}
with $\vec v(f,\vec e) \equiv \sqrt{2(2\pi f/m -1)} \vec e$.
$\Delta_{j}(f_n)$ is the velocity-weighted number fraction of DM waves normalized as $\sum_{j=x,y,z} \sum_n  \Delta_{j}(f_n) = 1$ since the left-handed side reproduces the definition of $\bar v^2$~\eqref{def_barv}.
The spectral shape of the spatial derivative of field value is determined by $\Delta_{j}(f_n)$.
Here, we consider typical cases of $\Delta_j$ where $\vec v_\odot$ is
perpendicular and proportional to $j$-direction.
After performing the angular integral, we obtain
\begin{align}   
    &\Delta_{\perp}(f_n)
    =
     \frac{v_\mathrm{vir}^2}{4(v_{\odot} ^2 + 3v_{\rm vir}^2/2)}
    \bigg[
        \text{erf}\left(\frac{v-v_\odot}{v_\mathrm{vir}}\right)
        +\text{erf}\left(\frac{v+v_\odot}{v_\mathrm{vir}}\right)
      \nonumber
    \\&\qquad
    -\frac{
    v_\mathrm{vir}                   e^{-\frac{(v+v_\odot)^2}{v_\mathrm{vir}^2}} 
    }{2\sqrt{\pi } v_\odot^3}
     \bigg(
         2v v_\odot 
         \left(e^{\frac{4 v v_\odot}{v_\mathrm{vir}^2}}+1\right)
\nonumber\\&\qquad
         +
          \left( 2v_\odot^2-v_\mathrm{vir}^2 \right)
          \left(e^{\frac{4 v v_\odot}{v_\mathrm{vir}^2}}-1\right)
     \bigg)
     \bigg]^{v(f_n+\Delta f/2)}_{v(f_n-\Delta f/2)}
     ,
     \label{eq_Delta_perp}
     \\
     &\Delta_{\parallel}(f_n)
    =
     \frac{(v_\odot^2+v_\mathrm{vir}^2/2)}{2(v_{\odot} ^2 + 3v_{\rm vir}^2/2) }
    \bigg[
        \text{erf}\left(\frac{v-v_\odot}{v_\mathrm{vir}}\right)
        +\text{erf}\left(\frac{v+v_\odot}{v_\mathrm{vir}}\right)
        \nonumber
    \\&\qquad
    -
    \frac{
    2v_\mathrm{vir}                   e^{-\frac{v^2+v_\odot^2}{v_\mathrm{vir}^2}} 
    }{\sqrt{\pi } v_\odot^3(2v_\odot^2+v_\mathrm{vir}^2)}
     \bigg(
         2v v_\odot
         (v_\odot^2 - v_\mathrm{vir}^2)
        \cosh\left(
        \frac{2v v_\odot}{v_\mathrm{vir}^2}
        \right)
\nonumber\\&\qquad        
         -
         (
         2 v^2 v_\odot^2
         +2 v_\odot^4
         +v_\mathrm{vir}^4
         )
         \sinh\left(
        \frac{2v v_\odot}{v_\mathrm{vir}^2}
        \right)
     \bigg)
     \bigg]^{v(f_n+\Delta f/2)}_{v(f_n-\Delta f/2)}
     .
     \label{eq_Delta_para}
\end{align}

For example, when we use the coordinate with $\vec v_\odot = (0,0,v_\odot)$, $\Delta_x=\Delta_y = \Delta_{\perp}$ and $\Delta_z = \Delta_{\parallel}$, respectively.
Note that $\Delta_{\perp}$ and $\Delta_{\parallel}$ are normalized so that 
\begin{align}
    \Delta_{\perp,\mathrm{tot}}
    &\equiv 
    \sum_n \Delta_{\perp}(f_n)
    \simeq
    0.19
    ,
\nonumber\\    
     \Delta_{\parallel,\mathrm{tot}}
    &\equiv 
    \sum_n \Delta_{\parallel}(f_n)
    \simeq
    0.62
    ,
    \label{eq_delta_pp_tot}
\end{align}
and $2\Delta_{\perp,\mathrm{tot}}+\Delta_{\parallel,\mathrm{tot}}=1$ holds. 

In general, the solar velocity points to a direction written by $\vec v= v_\odot(\cos\phi_\odot \sin\theta_\odot, \sin\phi_\odot \sin\theta_\odot,   \cos\theta_\odot )$.
A general formula of $\Delta_j$ for this direction is derived by rotating the coordinates, which is given by
\begin{align}
    \Delta_j 
    &=
    \begin{pmatrix}
    \Delta_\perp + \cos\phi_\odot^2 \sin\theta_\odot^2 (\Delta_\parallel-\Delta_\perp)
\\
    \Delta_\perp + \sin\phi_\odot^2 \sin\theta_\odot^2 (\Delta_\parallel-\Delta_\perp)
\\
    \Delta_\parallel -\sin^2\theta_\odot (\Delta_\parallel-\Delta_\perp)\quad\quad~
    \end{pmatrix}
    .
\end{align}

\begin{figure*}[tb]
  \centering
  \includegraphics[width=1.6\columnwidth]{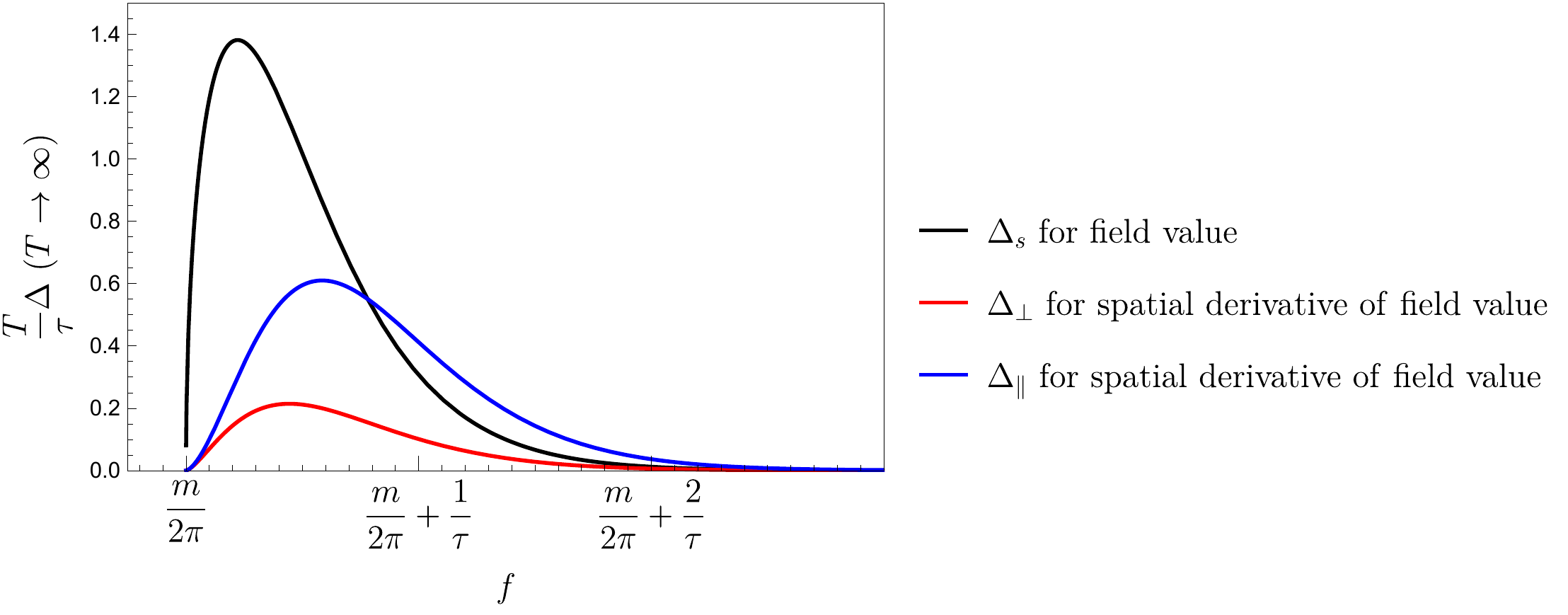}
  \caption{
  The deterministic part of normalized power spectrum in large $T$ limit.
  The vertical axis represents the spectral shape normalized by $T/\tau$. 
  A black, red, and blue lines represent the spectral shape for the field value~\eqref{eq_def_Delta_N}, the spatial derivative of the field value with perpendicular~\eqref{eq_Delta_perp} and parallel \eqref{eq_Delta_para} to the solar velocity, respectively.
  The actual amplitudes randomly fluctuate around them with the Rayleigh distribution.
  }
  \label{fig_powerspec_vector_compare}
\end{figure*}

$\Delta_{\perp}$ and $\Delta_{\parallel}$ have broader spectral shapes than $\Delta_s$ due to the velocity dependence. 
We compare $\Delta_{\perp}$ and $\Delta_{\parallel}$ with $\Delta_s$ in the large measurement time limit, $\int_{\Delta f} \df f \to 1/T$, which is shown in Fig.~\ref{fig_powerspec_vector_compare}.
All spectra have frequency ranges with $\mathcal O(1/\tau)$.
$\Delta_{\parallel}$ has larger amplitude and the broader spectral shape than $\Delta_{\perp}$, since the solar velocity $v_\odot$ is added to the DM velocity that contribute to the spatial derivative.
In Sec.~\ref{sec_sensitivity}, we estimate the upper limit of the coupling constant by using $\Delta_{\perp}$ and $\Delta_{\parallel}$.
Note that the above estimation holds for $T< 1~$day since the direction of velocity changes with the rotation of Earth, which is discussed in Sec.~\ref{sec_sensitivity_dark_photon}.

The above discussion is also applicable to a dark photon.
The massive dark photon has three independent bosonic degrees of freedom, and they have their own amplitudes and phases.
The Fourier transformation of the gauge field value is given by
\begin{align}
    \tilde A_k(f_n,\vec 0)
   &=
   \frac{T}{2}
   \sigma_A
    \sqrt{ \Delta_s(f_{n})}
    ~
   \frac{r_{k,n} }{{\sqrt{2}}}
   \exp(i\theta_{k,n})
    \label{eq_tildeA}
    ,
\end{align}
where the subscription ``$k$'' describes a quantity related to $A_k$. 
In the same way, the derivative of dark photon DM is given by
\begin{align}
    \nabla_j \tilde A_k
    (f_n,\vec x)
    \bigg|_{\vec x=0}
   &=
   \frac{T}{2}
   \sigma_A
   m \bar v
    \sqrt{ \Delta_j(f_n)  }
    ~
    \frac{r_{k,n} }{{\sqrt{2}}}
    \exp\left(
    i\theta_{k,n}
    \right)
    .
        \label{eq_nablatildeA}
\end{align}

\section{Probability distribution of experimental signals}
\label{sec_observation}

Before we estimate the stochastic effect on the DM searches, we revisit two examples of bosonic DM searches,
(i) axionlike particles observed by the polarization rotation of  laser light and (ii) dark photons observed via the displacement of massive equipment by a force sensor.
Here, we mainly consider the measurements through the gravitational-wave interferometer experiments while similar discussions are also applicable to other ultralight bosonic DM searches.
At first, we introduce an interaction of bosonic DM without specifying the  detection technique.
Then, we evaluate the probability distribution of the {DM} signals {that could be} observed by the interferometer search.

The axionlike particle, $a$, couples to the electromagnetic field through the Chern-Simons term,
\begin{align}
    \mathcal L_a =
    -
    \frac{1}{2}\partial_{\mu}a\partial_{\mu}a
    -\frac{1}{2}m^2 a^2
    +
    \frac{ g_a }{4} a(t) F_{{\rm EM},\mu\nu} \tilde F_{\rm EM}^{\mu\nu},
\end{align}
where $g_a$ is the axion-photon coupling constant, $F_{{\rm EM},\mu\nu}$ is electromagnetic field strength and $\tilde F_{\rm EM}^{\mu\nu}\equiv \epsilon^{\mu\nu\rho\sigma}F_{{\rm EM},\rho\sigma}/2$ is its Hodge dual with the Levi-Civita anti-symmetric tensor $\epsilon^{\mu\nu\rho\sigma}$.
Under an axion field background, a dispersion relation of photon is modified~\cite{Carroll:1989vb,Carroll:1998zi}, which is given for the left and right circular polarisation photon as $\omega_{L/R}^2 = k^2(1\mp g_a\dot a/k )$ with angular frequency $\omega_{L/R}$ and momentum $k$.
Thus, the background axion field changes the phase velocity of polarized photons, where the difference between left and right polarized photons is given by 
\begin{align}
    \delta c(t)
    \equiv
    \frac{g_a\dot a (t)}{2k}
    \label{eq_def_deltac}
    .
\end{align}

Next, dark photon is a U(1) massive gauge boson coupled to the $U(1)_D$ current $J_D^\mu$.
The massive dark photon with mass $m$  is described as
\begin{align}
    \mathcal L_A 
    &=
    -\frac{1}{4} F_{\mu\nu}F^{\mu\nu}
    +\frac{1}{2}m^2 A_\mu A^\mu
    -e \epsilon_DJ_D^\mu A_\mu,
\end{align}
where  $F_{\mu\nu}$ is the field strength of dark photon,
 and $ \epsilon_D$ is a dark photon coupling constant normalized to an electromagnetic one, $e$.
The background dark photon exerts the dark ``Lorentz force'' on  experimental equipment.
The dark electric force dominates the magnetic force since the momentum of DM is much smaller than its mass.
Then, the dark electric force acts on an object with charge $Q$ as $F(t,\vec x) =-e \epsilon_D  Q \dot{\vec A}(t,\vec x)$.
When there is no other force exerting the object, the displacement of the object is given by  
\begin{align}
    \delta {\vec x}(t,\vec x; q)
    =
    \int \df t^2~
    \frac{\vec F(t)}{M}
    =
    e \epsilon_D  
    \left(\frac{Q}{M}\right) 
    \frac{\dot{\vec A}(t,\vec x)}{m^2},
    \label{eq_displacement_dark_photon_interaction}
\end{align}
where $M$ is mass of the object.
In the last equality, we approximate that the field value oscillates with a frequency about $f\sim m/(2\pi)$. 
Note that the ratio of a charge to a mass, $(Q/M)$, depends on an ingredient.

In both cases of axion and dark photon, above interactions induce a periodic signal with the frequency $f \sim m/(2\pi)$.
The gravitational-wave interferometer experiments like aLIGO are sensitive to the signal with $10~{\rm Hz}\lesssim f_\mathrm{DM} \lesssim 10^3~{\rm Hz}$, which corresponds to the DM mass
\begin{align}
    m = 4.1\times 10^{-13} ~{\rm eV}\left( \frac{f_\mathrm{DM}}{10^2 {\rm Hz}} \right).
\end{align}
The corresponding coherent time and coherent length are given by
\begin{align}
    \tau = 0.3 ~\mathrm{day}~
    \frac{10^{-13}~\mathrm{eV}}{m}
   ,
    ~
    L_\phi =
    \frac{2\pi}{m \sqrt{\bar v^2}}
    \sim 
    10^{10}~\mathrm{m}
    \frac{10^{-13}~\mathrm{eV}}{m}
    .
\end{align}

\subsection{Axion DM}
\label{sec_axion}

In the interferometer experiments, the signal of axionlike particles would be measured as the polarization rotation of laser in interferometer arms.
In data analysis, we look for the DM signal in the frequency space.
By using Eqs~\eqref{eq_phi_fj_superposed} and \eqref{eq_dotPhi_formula}, the Fourier mode of the signal is given by
\begin{align}
     s_a(f_n)
    &\equiv
    \int_{-T/2}^{T/2} \df t e^{-2\pi i f_n t} \delta c(t) 
    \nonumber\\&
    \simeq
    g_a T
    \frac{ m \sigma_a }{4k}
    ~
    \sqrt{ \Delta_s(f_n)}
    ~
    \left[
    \frac{r_n }{{\sqrt{2}}}
    e^{i\theta_n}
    \right]
    ,
    \label{eq_signal_axions}
\end{align}
where we redefine the phase $\theta_n$ to include an unnecessary phase.

\subsection{Dark photon DM}
\label{sec_vector}

\begin{figure}[tb]
  \centering
  \includegraphics[width=0.8\columnwidth]{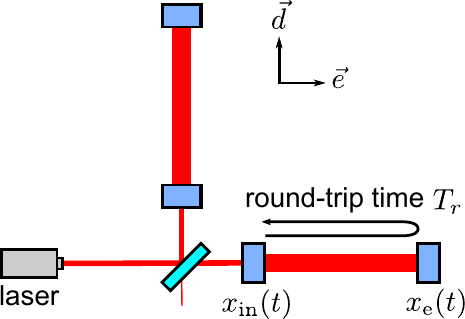}
  \caption{
  Configuration of laser interferometer arms.
  Two arms are parallel to $\vec d$ or $\vec e$.
  The laser makes a round trip between an input mirror at $x_\mathrm{in}$ and an end-test-mass mirror at $x_e$, respectively.
  }
  \label{fig_interferometer}
\end{figure}

$U(1)_D$ dark photon induces the dark electric force on a movable mirror.
In the interferometer search, we can detect it through the difference of light traveling time between two interferometer arms, which is given by~\cite{Morisaki:2020gui}
\begin{align}
    h(t) = \frac{\varphi(t,\vec e) -\varphi(t,\vec d)}{4 \pi \nu  L},
    \label{eq_sigmal_vector}
\end{align}
where $\nu$ is the frequency of the laser, $L$ is the length of interferometer arms. 
$\vec e$ and $\vec d$ are unit vectors directed to two interferometer arms, where $\vec e\cdot \vec d=0$ for LIGO, Virgo and KAGRA, and $\vec e\cdot \vec d=1/2$ for ET, DECIGO and LISA.
We describe the typical direction of the interferometer arms for LIGO-like experiment in Fig.~\ref{fig_interferometer}.
The phase $\varphi(t,\vec e)$ is gained through the reflection from the $\vec e$ arm, which is contributed by 
(i)  a finite-time traveling effect,
(ii) a spatial difference of the dark photon field value,
and 
(iii) a difference of the ingredient of two mirrors.
The first effect is pointed out for the dark photon search in Ref.~\cite{Morisaki:2020gui}.

The signals in the frequency space are written as 
\begin{align}
    s_A(f_n)
    &\equiv
    \int_{-T/2}^{T/2} \df t e^{- 2\pi i f_n t} h(t) 
    \simeq
    s_\mathrm{time}
    +s_\mathrm{space}
    +s_\mathrm{charge},
\end{align}
where each subscript corresponds to displacement of mirrors induced by the three effects.
Using stochastic representations of the dark photon~\eqref{eq_tildeA} and \eqref{eq_nablatildeA}, the signals are written by
(See App.~\ref{app_interferometer_dark_photon_signal} for a derivation.)
\begin{align}
	s_\mathrm{time}(f_n)
	&=
	\left(
	e \epsilon_D  T
	\left(\frac{Q}{M}\right)_\mathrm{in}
	\frac{\sigma_A   }{mL}
	\sin^2\left(\frac{mL}{2}\right)
	\sqrt{2}
	\right)
\nonumber\\&\times
	\sqrt{\Delta_s(f_n)}
	~
	\left[
	\frac{r_{n} }{{\sqrt{2}}}
	\exp(i\theta_{n})
	\right]
	,
	\label{eq_stime}
	\\
	s_\mathrm{space}(f_n)
	&=
	\left( 
	e \epsilon_D  T
	\left(\frac{Q}{M}\right)_\mathrm{in}
	\frac{ \sigma_A  \bar v}{2} 
	\right) 
\nonumber\\&\times
	\sqrt{\Delta_x(f_n)+\Delta_y(f_n)}
	~
	\left[
	\frac{r_{n} }{{\sqrt{2}}}
	\exp\left(
	i\theta_{n}
	\right)
	\right]
	\label{eq_sspace}
	,
	\\
	s_\mathrm{charge}(f_n)
	&=
	\left(
	e \epsilon_D T
	\left|
	\left(\frac{Q}{M}\right)_e-\left(\frac{Q}{M}\right)_\mathrm{in}
	\right|
	\frac{\sigma_A }{2Lm}
	\sqrt{2}
	\right) 
\nonumber\\&\times
	\sqrt{ \Delta_s(f_n)}
	~
	\left[
	\frac{r_{n} }{{\sqrt{2}}}
	\exp(i\theta_{n})
	\right]
	,
	\label{eq_scharge}
\end{align}
where we choose coordinates as ${\vec e}= (1,0,0)$ and ${\vec d}=(0,1,0)$ assuming the interferometer with two orthogonal arms like LIGO, Virgo, and KAGRA.
 $\left(Q/M\right)_\mathrm{in}$ and $\left(Q/M\right)_e$ are ratios of  charge to mass of the input and end-test-mass mirrors.
$\theta_{n}$ follows a uniform distribution over $[0,2\pi]$, and  $r_{n}$ follows a standard Rayleigh distribution.
The spectral shape of $s_\mathrm{time}$ and $s_\mathrm{charge}$ follows $\Delta_s$, which is the same to the axion signal~\eqref{eq_signal_axions}.
On the other hand, the spectral shape of $s_\mathrm{space}$ is determined by $\Delta_x$ and $\Delta_y$, which depends on the direction of the solar velocity $\vec v_\odot$ relative to the interferometer arms.
Here, we consider two typical directions: (i) a conservative direction with $\vec v_\odot = (0,0,v_\odot)$ leads to $\Delta_x = \Delta_y = \Delta_\perp$, and (ii) an optimal direction with $\vec v_\odot = (v_\odot,0,0)$ leads to $\Delta_x=\Delta_\parallel,~\Delta_y= \Delta_\perp$.  
In other words, the Sun moves in a direction perpendicular to the interferometer arms in the conservative case and parallel to one of the interferometer arms in the optimal case.
Compared to $\Delta_s$, $\Delta_x+\Delta_y$ has broader distribution on the frequency space. 
We discuss constraints in the conservative and optimal cases in Sec.~\ref{sec_velocity_independent}.

\begin{figure}[tb]
  \centering
  \includegraphics[width=1.\columnwidth]{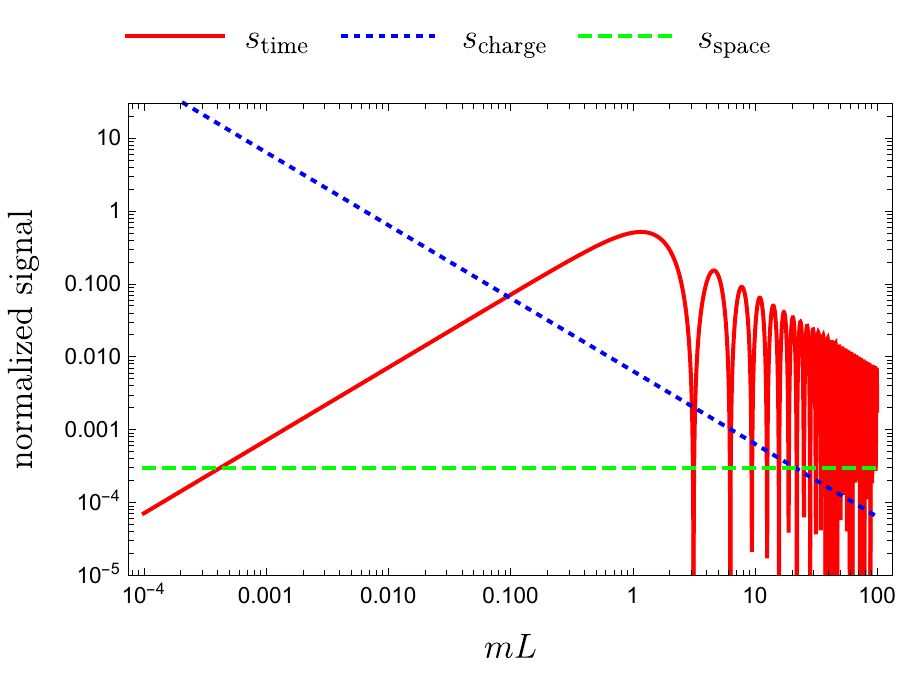}
  \caption{
  Normalized signals of a dark photon, $s_X/(e \epsilon_D  \sigma_A T/m_n)$, for $X=$ time (red), charge (blue dotted) and space (green dashed). These $s_X$ are given in Eqs.~\eqref{eq_stime}, \eqref{eq_sspace} and \eqref{eq_scharge}. 
    We adopt typical parameters, $m_n \left(Q/M\right)_\mathrm{in}=0.501$, $m_n |\left(Q/M\right)_e- \left(Q/M\right)_\mathrm{in}| = 0.51-0.501$, $|\vec e-\vec d|=\sqrt{2}$ and  $\bar v\simeq 1.2\times 10^{-3}$.
    We ignore the stochastic effect  by setting  $(r_n/\sqrt{2})\exp(i\theta_n)=1$ and fix the spectral amplitudes with $\Delta_s=1$ and $\Delta_x+\Delta_y=1$ in this figure.
  }
  \label{fig_signal_vector_compare}
\end{figure}

We present the normalized signals in Fig.~\ref{fig_signal_vector_compare} to compare the strength of three signals.
We take the typical parameters, 
$\bar v\simeq 1.2\times 10^{-3}$, 
$m_n \left(Q/M\right)_\mathrm{in}=0.501$, and
$m_n |\left(Q/M\right)_e- \left(Q/M\right)_\mathrm{in}| = 0.51-0.501$ 
with the neutron mass $m_n$.
The stochastic effect and spectral shape are fixed in this figure by setting $2^{-1/2}r_n\exp(i\theta_n)\to1$, $\Delta_s\to1$, and $\Delta_x+\Delta_y\to1$.
First, the signal from the finite-time-traveling effect,  $s_\mathrm{time}$, is prominent for $mL\gtrsim 1$, while it is suppressed for $mL\ll 1$ due to a short traveling time. Second, the signal from a charge difference of mirrors, $s_\mathrm{charge}$, depends on the coupling of dark photon and ingredients of mirrors. For the KAGRA-like interferometer, the  mirrors are made of fused silica and sapphire.
The $B$-$L$ charge of fused silica and sapphire are $(Q/M)=0.501/m_n$ and $0.51/m_n$. 
While $s_\mathrm{charge}$ is suppressed by $m_n|\left(Q/M\right)_e- \left(Q/M\right)_\mathrm{in}|\simeq 10^{-2}$, $s_\mathrm{charge}$ is more sensitive for dark photon with a small mass than the other two signals.
It is because $s_\mathrm{charge}$ is induced by $\dot {\vec A}$ while other two signals are induced by the derivative of $\dot{\vec A}$. 
Third, the signal from a spatial difference of field value, $s_\mathrm{space}$, is effective for $mL\gg 1$. 
This signal is highly suppressed by a small DM velocity by a factor $\bar v \sim 10^{-3}$.
Note that the expression is valid only for $L m \bar v \ll1$, {because of the linear approximation made in Eq.~\eqref{eq:deriveapp}}.
Considering Fig.~\ref{fig_powerspec_vector_compare}, all the signals are important when we search for DM with various masses.
In the following section, we estimate the sensitivity to these signals.

\section{Estimation of an upper bound on a coupling constant}
\label{sec_sensitivity}

In this section, we derive the likelihood function to constrain the DM coupling constant for a given experimental data, taking into account the stochastic nature of DM signal 
derived in Sec.~\ref{sec_Stochastic_DM}.
The experimental data consists of the instrumental noises and the DM signal.
For the signal of light DM, the random fluctuations come from not only the noises but also the stochastic DM signal.
The stochasticity of the DM signal affects the {resultant} likelihood function.
We confirm the derived likelihood function by using numerical simulation.

\subsection{Likelihood function of data}
\label{sec_likelihood}

Interferometer experiments measure a time-series data on laser power.
Here, we assume a measurement time $T$ and investigate the Fourier-transformed data to extract the periodic signal of the bosonic DM.
The measured data is written as $\mathcal N(f_n)+s(f_n)$, where $\mathcal N(f_n)$ is instrumental noise, $s(f_n)$ is a signal in the frequency space, and $f_n$ is the $n$-th frequency bin with a finite frequency resolution $\Delta f= 1/T$.
We assume that the instrumental noise follows the Gaussian distribution characterized by its power spectral density, $S_{\rm noise}(f)$.

We define normalized data at a frequency $f_n$ around the oscillation frequency $f_\mathrm{DM} = m/(2\pi)$ as
\begin{align}
   \rho_n
   &\equiv 
   \frac{
   4  |\mathcal N(f_n) + s(f_n)|^2
   }{T S_{\rm noise}(f_n)}
   =
   |\hat {\mathcal{N}}_n + \lambda_n r_n e^{i\theta_n} |^2
    \label{eq_def_d0},
    \\
    \hat {\mathcal{N}}_n
    &\equiv
    \frac{  2{\mathcal N}(f_n) }{
    \sqrt{ T S_{\rm noise}(f_n)  }  }
    ,
\end{align}
where $r_n$ follows a standard Rayleigh distribution, and $\theta_n$ follows the uniform distribution.
The normalized noise, $\hat {\mathcal{N}}_n$, is a complex stochastic variable following the Gaussian distribution,
\begin{align}
    P_{\rm noise}(\hat {\mathcal{N}}_n)~ {\rm d}^2 \hat {\mathcal{N}}_n 
    \equiv \frac{1}{{2}\pi} \exp(-| \hat {\mathcal{N}}_n |^2/{2})
    ~\df{ \rm Re}[\hat {\mathcal{N}}_n] \df {\rm Im}[\hat {\mathcal{N}}_n] .
    \label{eq_noise_gauss}
\end{align}
$\lambda_n$ is the amplitude of the DM signal normalized by the noise power spectrum, which is decomposed by the frequency independent and dependent parts,
\begin{align}
    \lambda_n
     &\equiv 
    \bar \lambda_X \sqrt{\Delta_X(f_n)},
\end{align}
where $X$ is the label of signal type.
According to the discussion in Sec.~\ref{sec_observation}, the explicit formula for an axion signal~\eqref{eq_signal_axions}, and dark photon signals~\eqref{eq_stime}-\eqref{eq_sspace} for a interferometer with two orthogonal arms like LIGO are written by
\begin{align}
    \bar \lambda_\mathrm{axion}
     &=
     \frac{  T  }{ \sqrt{T S_{\rm noise}}}
     g_a \sqrt{\frac{\rho_{\rm DM}}{m^2}}
    \frac{ m }{2k}
,\quad
\nonumber
\\
    \bar \lambda_\mathrm{time}
     &=
    \epsilon_D e 
    \frac{2  T  }{ \sqrt{T S_{\rm noise}}}
    \sqrt{\frac{2\rho_{\rm DM}}{3m^2}  }
    \frac{(Q/M)_\mathrm{in}   }{mL}
    \sin^2\left(\frac{mL}{2}\right)
,\quad
\nonumber
\\
    \bar \lambda_\mathrm{space}
     &=
    \epsilon_D e 
    \frac{2  T  }{ \sqrt{T S_{\rm noise}}}
    \sqrt{\frac{2\rho_{\rm DM}}{3m^2}  }
    \frac{ (Q/M)_\mathrm{in}   \bar v}{2{\sqrt{2}}} 
,\quad
\nonumber
\\
    \bar \lambda_\mathrm{charge}
     &=
    \epsilon_D e 
    \frac{2  T  }{ \sqrt{T S_{\rm noise}}}
    \sqrt{\frac{2\rho_{\rm DM}}{3m^2}  }
    \frac{|(Q/M)_e-(Q/M)_\mathrm{in}|}{2Lm}
.
\label{eq_def_barlambda}
\end{align}
See App.~\ref{app_interferometer_dark_photon_signal} for a signals in a general interferometer configuration.
$\Delta_X(f_n)$ denotes {the deterministic part of} the spectral shape of the signals 
and is given by 
\begin{align}
    \Delta_X(f_n)= 
    \begin{cases}
          \Delta_s(f_n)
          \\
           \Delta_x(f_n)+\Delta_y(f_n)
    \end{cases}
    ,
    \label{eq_choice_Delta}
\end{align}
where the above case is for the velocity-independent signals ($X=$axion,~time,~charge), the below case is for the velocity-dependent signal ($X=$space), and $\Delta_s$ and $\Delta_j$ are given in Eqs.~\eqref{eq_def_Delta_N} and ~\eqref{eq_def_Delta_v}, respectively.
The signals with $X=$ time and space could be comparable in a gravitational wave observatory. 
Then, the combined signal is given by a root of squared sum of both signals since a phase between $s_\mathrm{time}$ and $s_\mathrm{space}$ is different by a factor $i$.


The likelihood for data $\rho_n$  
with  
a given DM signal $\lambda_n r_n$ is expressed as
\begin{align}
    &\mathcal L(\rho_n|\lambda_nr_n)
\nonumber\\&\equiv 
    \int{\rm d}^2 \hat {\mathcal{N}}_n
    ~ P_\mathrm{noise}(\hat {\mathcal{N}}_n)
    \delta\left(
        \rho_n 
        - \left| \hat {\mathcal{N}}_n + \lambda_nr_n ~e^{i\theta_n} \right|^2 
    \right)
    \nonumber\\&=
    {\frac{1}{2}}
    \exp\left(
    -\frac{\rho_n+(\lambda_nr_n)^2}{{2}}
    \right)
    I_0(
    \sqrt{\rho_n}\lambda_nr_n
    ),
    \label{eq_def_L}
\end{align}
where $I_0(x)$ is the modified Bessel function of the first kind.
Note that the signal phase $e^{i\theta_n}$ can be absorbed by the noise phase and its dependence vanishes after the integration of the noise.
The signal fluctuates due to the stochastic nature of the field value, and the random variable $r_n$ represents the stochastic effect, which follows the Rayleigh distribution. Its squared average is $\int  {\rm  d}r_n
    P_R (r_n) ~r_n^2 = 2$.
We call ${\mathcal L}(\rho_n|\lambda_n\sqrt{2})$ a likelihood in a deterministic case~\cite{Centers:2019dyn}, in which we neglect the Rayleigh distribution of the field value and replace $r_n$ by its RMS value $\sqrt{2}$.
We marginalize the likelihood over $r_n$ as
\begin{align}
    \overline{\mathcal L}(\rho_n|\lambda_n)
    & \equiv 
    \int  {\rm  d}r_n~
    P_R (r_n)
     {\mathcal L}\left (\rho_n \big | \lambda_n r_n \right)
\nonumber\\&=
     \frac{1}{{2}(1+\lambda_n^2)} \exp\left( 
     \frac{-\rho_n}{{2}(1+\lambda_n^2)}
     \right)
     .
     \label{eq_barL_single}
\end{align}
We call $\overline{\mathcal L}(\rho_n|\lambda_n)$ the likelihood in a stochastic case in contrast with a deterministic case, ${\mathcal L}(\rho_n|\lambda_n\sqrt{2})$.
The marginalization over $r_n$ broadens the distribution of $\overline{\mathcal L}(\rho_n|\lambda_n)$ due to the field fluctuation unlike ${\mathcal L}(\rho_n|\lambda_n\sqrt{2})$, which will weaken the constraint on the DM coupling constant~\cite{Centers:2019dyn}.

Since the DM signal is extended in the frequency space around $f_\mathrm{DM} = m/(2\pi)$, we take the summation of the power spectrum over the frequency range, 
\begin{align}
    \rho \equiv  
    \sum_{f_\mathrm{DM}< f_n <f_\mathrm{DM}(1+\kappa \bar v^2)}
    \rho_n
    ,
    \label{eq:freq_range}
\end{align}
where {we introduced a new parameter} $\kappa$ representing the {frequency} range to include the tail of DM distribution.
In this paper, we will use $\kappa \simeq 1.69$ for the velocity-independent signals and $\kappa \sim 2$ for the velocity-dependent signals so that the ranges cover 99\% of DM signals.
We discuss the choice of $\kappa$ in App.~\ref{sec_approx_measurmetn}.
Considering the finite frequency resolution $\Delta f = 1/T$, the number of bins in this range is estimated as 
\begin{align}
    N_\mathrm{bin} 
    = 
      \left\lceil
    \frac{ \kappa \bar v^2f_\mathrm{DM}}{ \Delta f}
    \right\rceil
    =
    \left\lceil
    \kappa 
    \frac{T}{\tau }
    \right\rceil
    ,
    \label{eq_num_bin}
\end{align}
where $\lceil x\rceil$ represents the minimum integer larger than $x$.

The likelihoods of $\rho$ in a deterministic and a stochastic case are estimated in App.~\ref{app_derive_likelihood}, and results are
\begin{align}
 \mathcal{L}(\rho | \{\lambda_n\}) 
&\equiv
    \int
	\left(
	\prod_n^{N_\mathrm{bin}}
	{\rm d} \rho_n~
	{\mathcal L} (\rho_n|\sqrt{2}\lambda_n)
	\right)
	\delta\left(\rho - \sum_n^{N_\mathrm{bin}} \rho_n  \right)
\nonumber\\&=
    \frac{e^{-(\rho+\Lambda)/2}}{2}
    \left(\frac{\rho}{\Lambda }\right)^{\frac{{N_\mathrm{bin}}-1}{2}}
    I_{{N_\mathrm{bin}}-1} (\sqrt{\Lambda\rho}),
    \label{eq_likelihood_deterministic}
\\
	\overline{\mathcal L}(\rho|\{\lambda_n\})
&\equiv
	\int
	\left(
	\prod_{l }^{N_\mathrm{bin}}
	{\rm d} \rho_n~
	\overline{\mathcal L} (\rho_n|\lambda_n)
	\right)
	\delta\left(\rho - \sum_n^{N_\mathrm{bin}} \rho_n  \right)
\nonumber\\&=
    \sum_n^{N_\mathrm{bin}}  
     \frac{
    w_n
     }{{2}(1+\lambda_n^2)}
      \exp\left( 
     - \frac{\rho}{{2}(1+\lambda_n^2)}
     \right)
	,
	\label{eq_label_likelihood}
\\
    \Lambda 
    &\equiv 2\sum_n^{N_\mathrm{bin}}(\lambda_n)^2
\quad,\quad
	w_n
	\equiv
	\prod_{n'(\neq n)}^{N_\mathrm{bin}}
	\frac{1+\lambda_n^2}{\lambda_n^2-\lambda_{n'}^2}
	,
\end{align}
where we assume $\lambda_n\neq \lambda_n'$ for all $n\neq n'$ in a stochastic case, and a formula with $\lambda_n= \lambda_n'$ is shown in App.~\ref{app_derive_likelihood}.
$I_{n-1}$ is the modified Bessel function of the first kind. 
Note that ${\mathcal L}(\rho|\{\lambda_n\})$ follows the noncentral chi-square distribution when the number of degrees of freedom {is} $2{N_\mathrm{bin}}$, and the noncentrality parameter $\Lambda$.
The numerical estimate of $\overline{\mathcal L}$ requires a large computational cost for a large $N_\mathrm{bin}$ due to  huge $w_n$'s. 
In this case, one can use an approximate formula in Eq.~\eqref{eq_approx_likelihood} to reduce the computational cost.

We see the difference between the stochastic and deterministic cases in the following sections.
In Ref.~\cite{Centers:2019dyn}, the authors found that field fluctuations loosen the upper bound of the coupling constant by a factor of about three in the case of $N_\mathrm{bin} =1$, where they compared the deterministic likelihood~\eqref{eq_def_L} and the stochastic one~\eqref{eq_barL_single}.
In Sec.~\ref{sec_experiment}, we revisit this point and extend their analysis to $N_\mathrm{bin} >1$ by using the likelihood of $\rho$ in Eqs.~\eqref{eq_likelihood_deterministic} and \eqref{eq_label_likelihood}.

By using the above likelihood, we discuss an upper bound on the coupling constant by frequentist's method.
When we conduct an experiment,
obtain a observed data $\rho_\mathrm{obs}$,
and do not find any signal of DM, 
we can set the upper bound on the DM coupling constant based on the likelihood function. 
We regard the power spectrum of background noise $S_\mathrm{noise}(f)$ as constant in a frequency range $f\in [f_\mathrm{DM},f_\mathrm{DM}(1+\kappa \bar v^2)]$  since the typical time scale of noise fluctuations is much larger than the coherence time.
For a given observed data $\rho_\mathrm{obs}$, the upper bound on the DM coupling constant is determined in the following way. 
We seek to put the upper bound of $\bar \lambda_X$ defined in Eq.~\eqref{eq_def_barlambda}, which is easily translated into that of the coupling constant.
We determine the upper bound on $\bar \lambda_X$ so that the false exclusion of the true signal occurs by a probability smaller than $1-\beta$.
In this paper, we take $1-\beta = 0.05$ as an example. 
The upper bound, $ \bar \lambda_{\rm up}$, is determined by the following integration:
\begin{align}
    1-\beta = 
	\int_0^{\rho_{\rm obs}}{\rm d} \rho
	~
	\overline{\mathcal L} 
	\left(
	    \rho
	    \big |
    	\left\{
    	\lambda_n
    	= 
    	\bar\lambda_\mathrm{up}\sqrt{\Delta_X(f_n)}
    	\right\}
    \right).
	\label{eq_def_lambdaup}
\end{align}
The right-handed side of Eq.~\eqref{eq_def_lambdaup} decreases as $\bar \lambda_\mathrm{up}$ increases. 
Thus, the smaller $1-\beta$ leads to the looser upper bound on the coupling constant.

\subsection{Simulation of dark matter signals}
\label{sec_simulation}

We performed numerical simulations to validate our algorithm of calculating the upper bound on a coupling constant.
We generated hundreds of simulated data containing dark matter signal, and applied our algorithm to the simulated data.
If the algorithm works properly, the upper bound with a confidence level of $p$ should be larger than the true value for a fraction $p$ of simulations.
For example, the $90\%$ upper bounds should be larger than the true value for $90\%$ of simulations.

We consider $U(1)_B$ dark photon and simulated dark-photon signal observed by an interferometric gravitational-wave detector.
For each simulation, a dark photon field was calculated as the sum of $10^4$ partial waves, whose velocities were drawn from the velocity distribution of the standard halo model, constant phases were drawn uniformly between $0$ and $2 \pi$, and polarization vectors were randomly chosen among $3$ unit vectors pointing in the $x$, $y$, and $z$ directions.
The signal was calculated from the generated dark photon field, and injected into Gaussian noise colored by the design sensitivity of advanced LIGO \cite{aligosensitivity}.
We assume that the two arms of the detector point in the $x$ and $y$ directions, and the velocity of the Sun points in the $z$ direction.
For each simulated data, the detection statistic $\rho$ was calculated, and the upper bound of $\epsilon_B$ was obtained.
For the choice of frequency bins to sum over, $\kappa=2$ was applied.

We consider two fiducial values of the signal frequency $f_\mathrm{DM}$: $20\,\mathrm{Hz}$ and $100\,\mathrm{Hz}$.
$s_{\mathrm{space}}$ and $s_{\mathrm{time}}$ are comparable in the former case as $\bar \lambda_\mathrm{space} / \bar \lambda_\mathrm{time} \simeq 1$, while $s_{\mathrm{time}}$ is dominant in the latter case as $\bar \lambda_\mathrm{space} / \bar \lambda_\mathrm{time} \simeq 0.2$.
For each frequency, we generated two sets of data of different duration, $T=\tau / 10$ and $T=5 \tau$.
The stochastic effect of signals is expected to be significant in the former case, while it is suppressed in the latter case.

We simulate the case where the signal and the noise are comparable since this paper aims to put the upper bound of the coupling constant.
For this purpose, we tuned the true value of $\epsilon_B$ so that the expected signal to noise ratio is moderate for each measurement time. 
For $T=\tau / 10$, only a single frequency bin contributes to $\rho$.
Given $n$ denotes the index of that frequency bin, $\epsilon_B$ was determined so that $\left<\lambda^2_n r^2_n\right>=2 \lambda^2_n = 5$, which leads to $\left<\rho\right>=7$, where the braket represents the expectation value for the stochastic realizations.
The sensitivity to signal is improved for longer integration time, and the detectable value of $\epsilon$ scales with $1 / \min \left[T^{\frac{1}{2}}, (\tau T)^{\frac{1}{4}}  \right]$~\cite{Budker:2013hfa}.
Thus, the detectable value of $\epsilon_B$ for $T=5 \tau$ is $4.7$ times smaller than that for $T = \tau / 10$, and the true value of $\epsilon_B$ for the longer duration was set to be $4.7$ times smaller than that for $T=\tau / 10$.
More concretely, we considered the following 4 pairs of frequency, duration, and coupling constant,
\begin{equation}
\begin{aligned}
    (f_\mathrm{DM}, T, \epsilon_B) 
    = & 
    (20\,\mathrm{Hz},~3.6\times10^3\,\mathrm{s},~1.1\times10^{-22}), 
   \nonumber \\&
    (20\,\mathrm{Hz},~1.8\times10^5\,\mathrm{s},~2.3\times10^{-23}),
\nonumber \\&
    (100\,\mathrm{Hz},~7.1\times10^2\,\mathrm{s},~5.1\times10^{-23}), 
\nonumber \\&
    (100\,\mathrm{Hz},~3.6\times10^4\,\mathrm{s},~1.1\times10^{-23}).
    \end{aligned}
\end{equation}
For each pair, we generated $400$ realizations of data.

\begin{figure*}
    \begin{center}
        \centering
        \includegraphics[width=0.8\columnwidth]{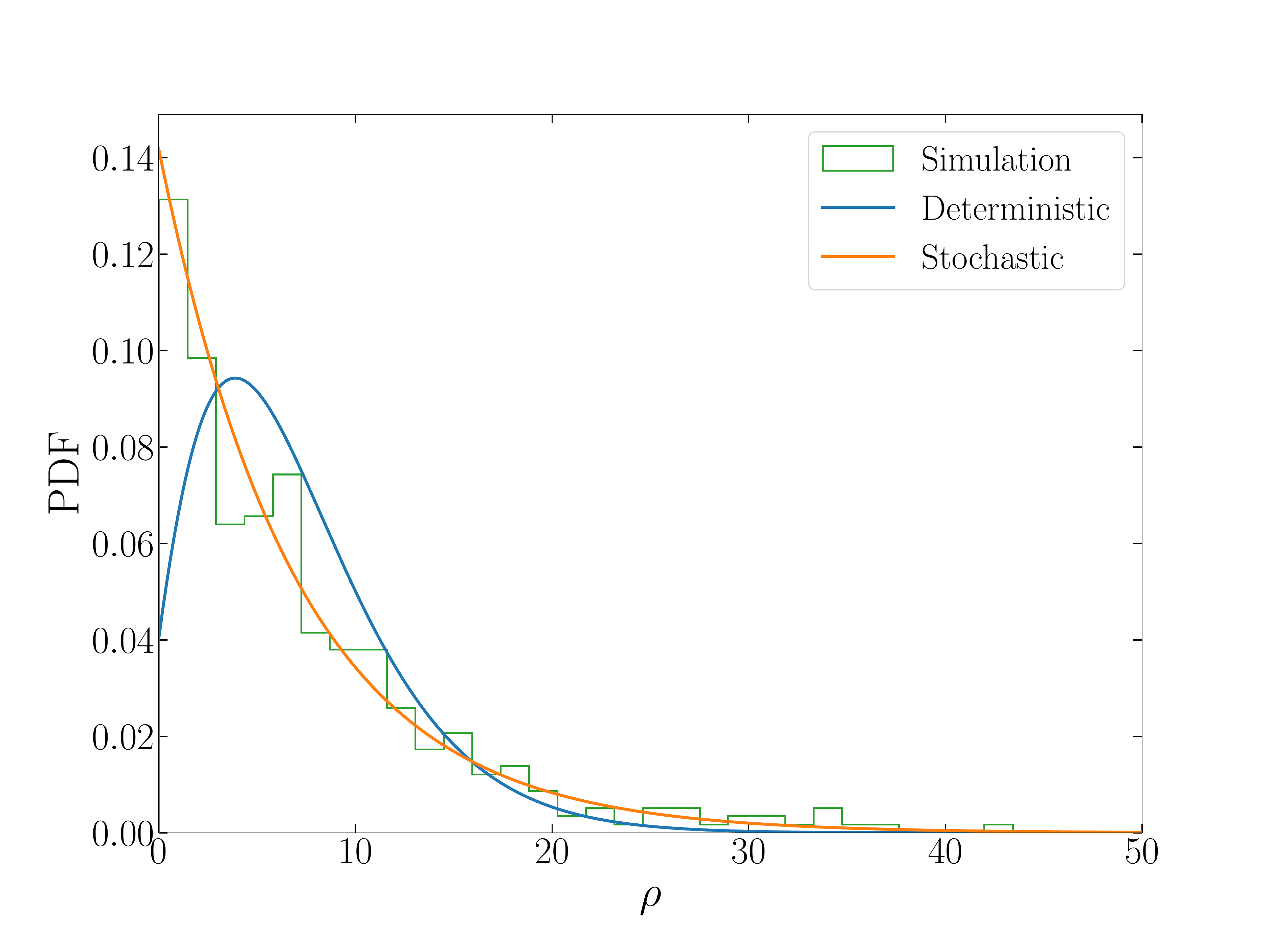}
        \includegraphics[width=0.8\columnwidth]{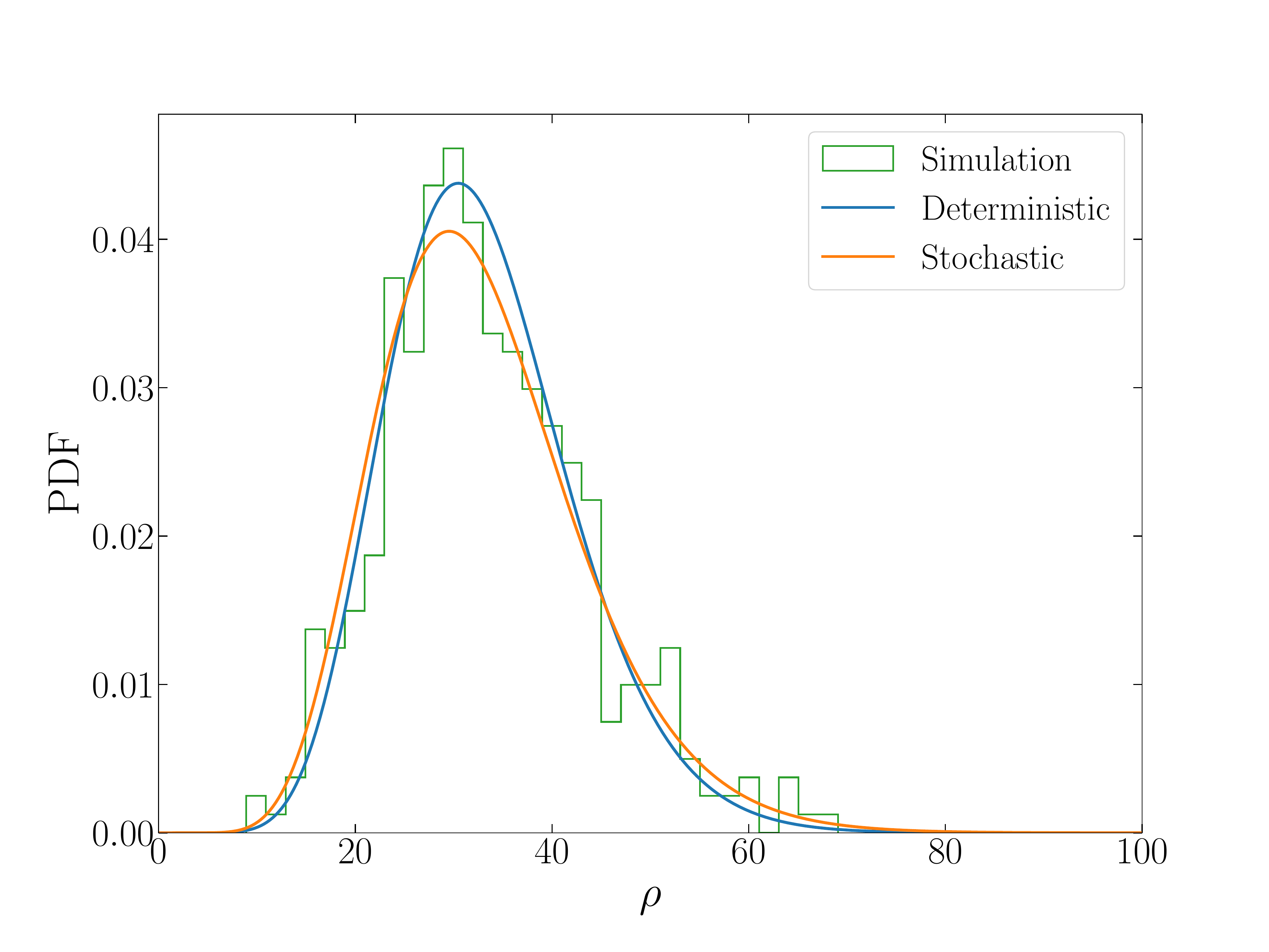}
        \includegraphics[width=0.8\columnwidth]{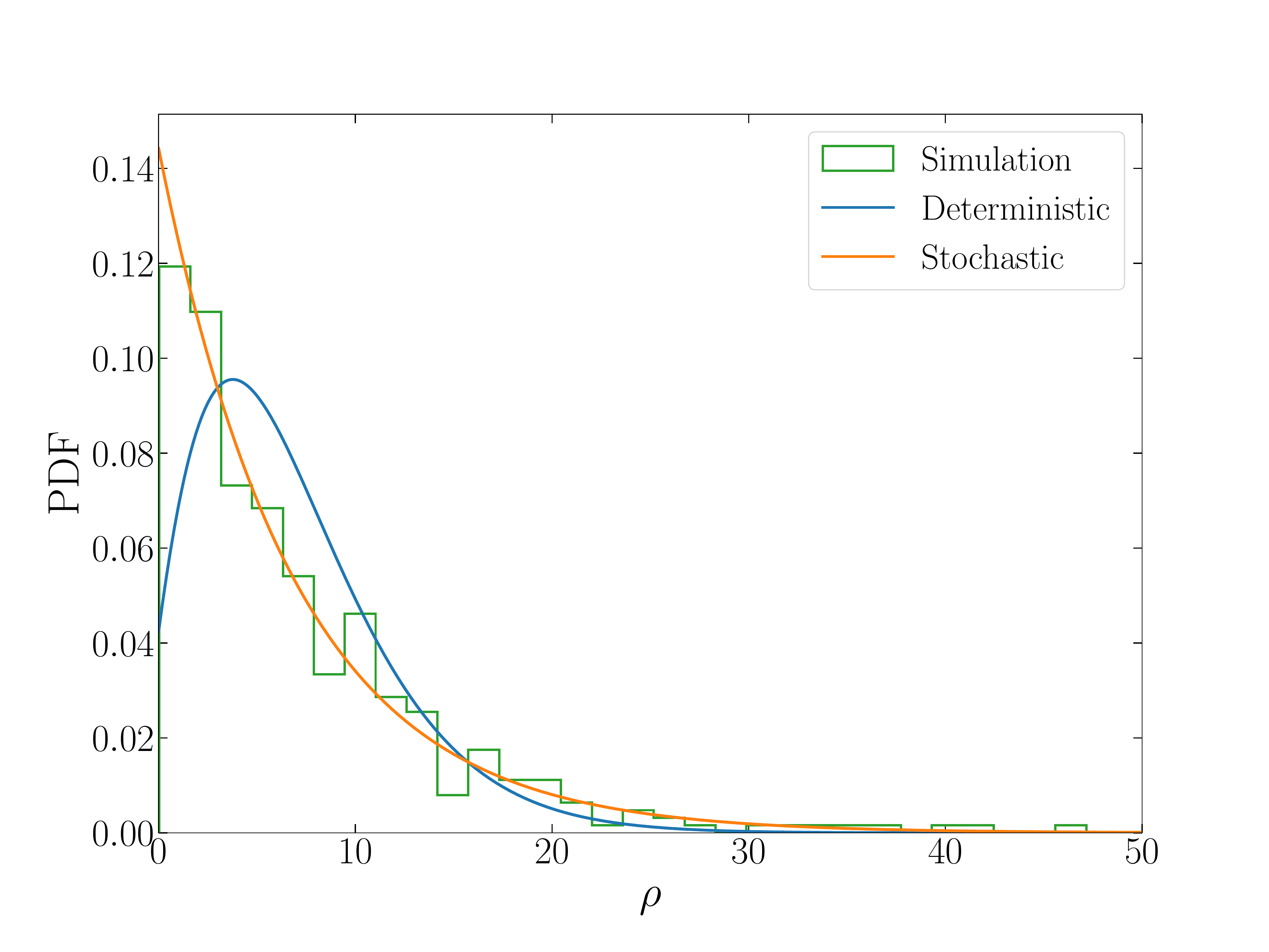}
        \includegraphics[width=0.8\columnwidth]{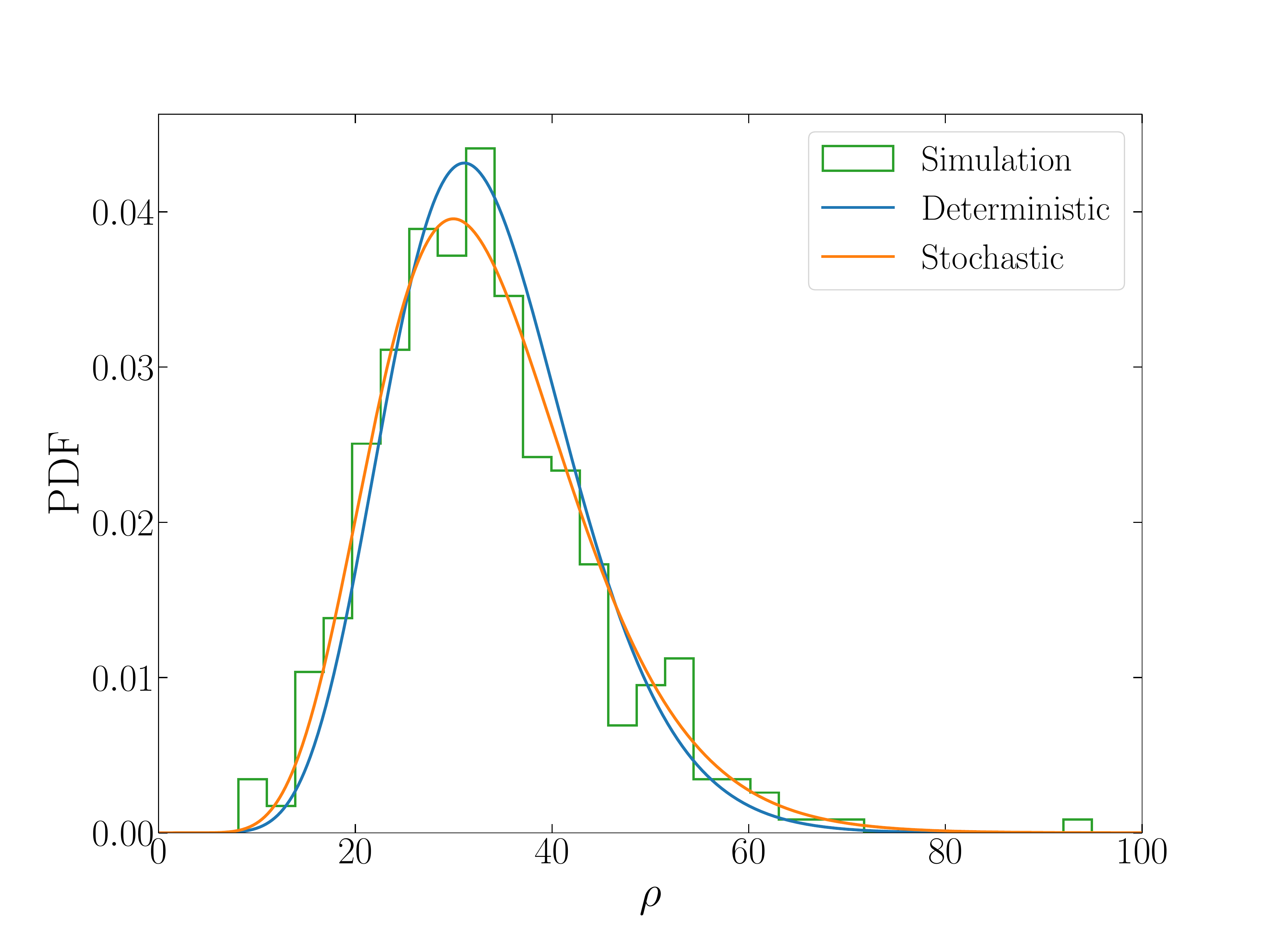}
        \caption{The probability distribution of $\rho$ obtained from $400$ simulated data in comparison with the deterministic (blue) and stochastic (orange) models. The upper left panel is for $(f_\mathrm{DM}, T, \epsilon_B) = (20\,\mathrm{Hz},~3.6\times10^3\,\mathrm{s},~1.1\times10^{-22})$, the upper right for $(20\,\mathrm{Hz},~1.8\times10^5\,\mathrm{s},~2.3\times10^{-23})$, the lower left for $(100\,\mathrm{Hz},~7.1\times10^2\,\mathrm{s},~5.1\times10^{-23})$, and the lower right for $(100\,\mathrm{Hz},~3.6\times10^4\,\mathrm{s},~1.1\times10^{-23})$.}
        \label{fig_snr}
    \end{center}
\end{figure*}

Figure~\ref{fig_snr} shows the histograms of $\rho$ obtained from simulated data.
For comparison, each panel shows the model probability distribution in the deterministic and stochastic cases, which are given in Eq.~\eqref{eq_likelihood_deterministic} and Eq.~\eqref{eq_label_likelihood} respectively.
In either value of frequency, the deterministic model fails to fit the observed distribution for $T=\tau / 10$, while the stochastic model fits it well.
The $p$--values of the Kolmogorov-Smirnov tests for $f_\mathrm{DM}=20\,\mathrm{Hz}~(100\,\mathrm{Hz})$ are $2.6 \times 10^{-6}~(1.6\times10^{-6})$ and $0.32~(0.75)$ for the deterministic and stochastic models respectively.
On the other hand, the stochastic model asymptotes to the deterministic model for $T=5 \tau$, and either model fits the observed distribution well.
The $p$--values of the Kolmogorov-Smirnov tests for $f_\mathrm{DM}=20\,\mathrm{Hz}~(100\,\mathrm{Hz})$ are $0.56~(0.13)$ and $0.58~(0.47)$ for the deterministic and stochastic models respectively.
Thus, the stochastic effect is negligible in a long measurement time.
It is because we choose the small coupling constant in the simulation for $T=5 \tau$ and the stochastic effect becomes subdominant compared to the experimental noise.
This setup of simulation corresponds to realistic experiments since experiments with a long measurement time inevitably search for such a small coupling constant to put an upper bound.
Moreover, both distributions converge to the Gaussian distribution in a large measurement time.
In Sec.~\ref{sec_experiment}, we will show that the stochastic nature of DM becomes negligible to estimate the upper bound of the coupling constant in a long measurement time.

To assess the impact of each of $s_{\mathrm{space}}$ and $s_{\mathrm{time}}$ for $f_\mathrm{DM}=20\,\mathrm{Hz}$, we also compute $p$--values in comparison with the stochastic model in which each contribution is turned off.
The $p$--values are reduced to $3.5 \times 10^{-6}$ and $6.7 \times 10^{-8}$ for $T=\tau / 10$ and $T = 5 \tau$ respectively if $s_{\mathrm{space}}$ is turned off, and $7.2 \times 10^{-34}$ and $4.9 \times 10^{-49}$ if $s_{\mathrm{time}}$ is turned off.
The results show that both contributions need to be taken into account to fit the observed distribution, and our model accurately incorporates both of them.

\begin{figure*}
    \begin{center}
        \centering
        \includegraphics[width=0.8\columnwidth]{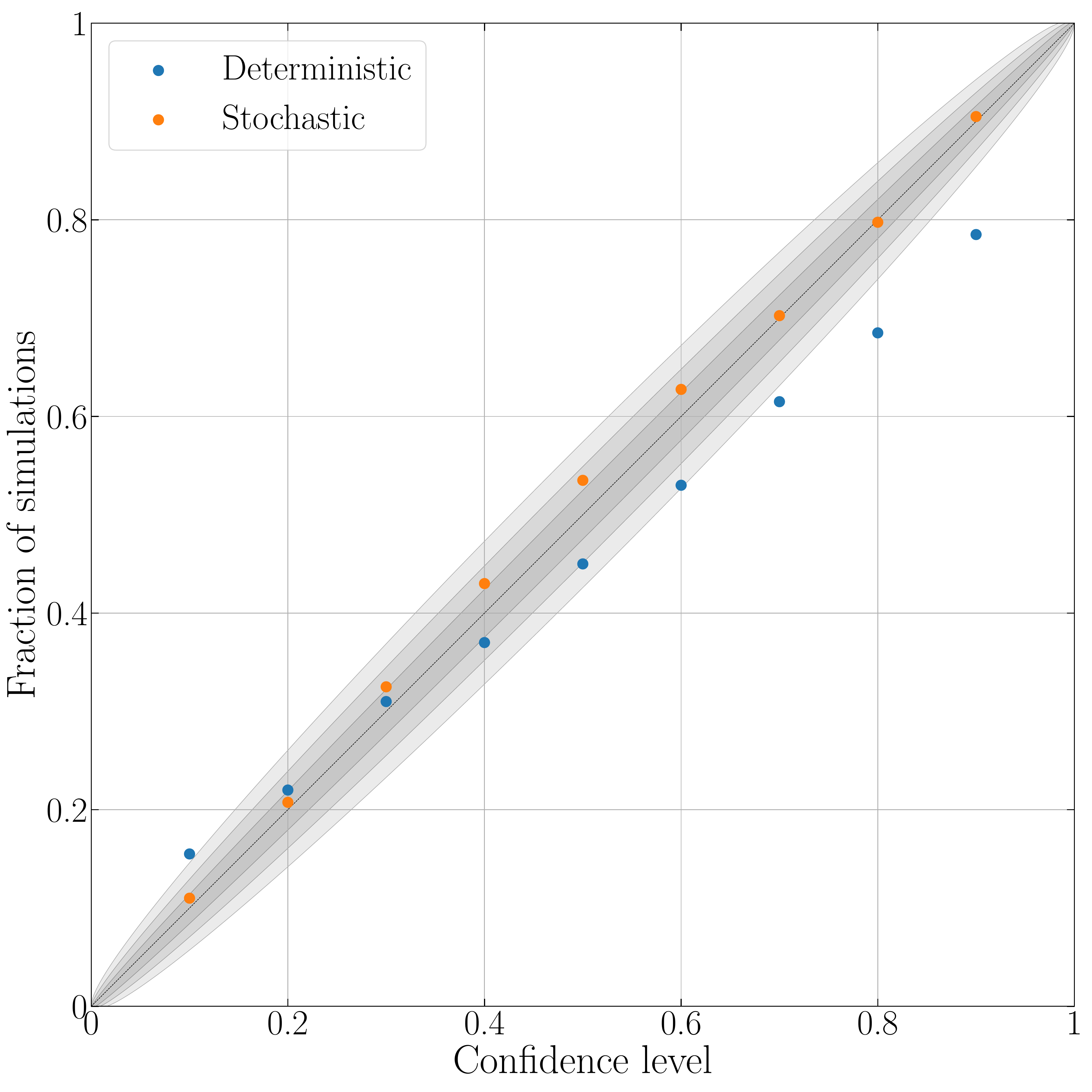}
        \includegraphics[width=0.8\columnwidth]{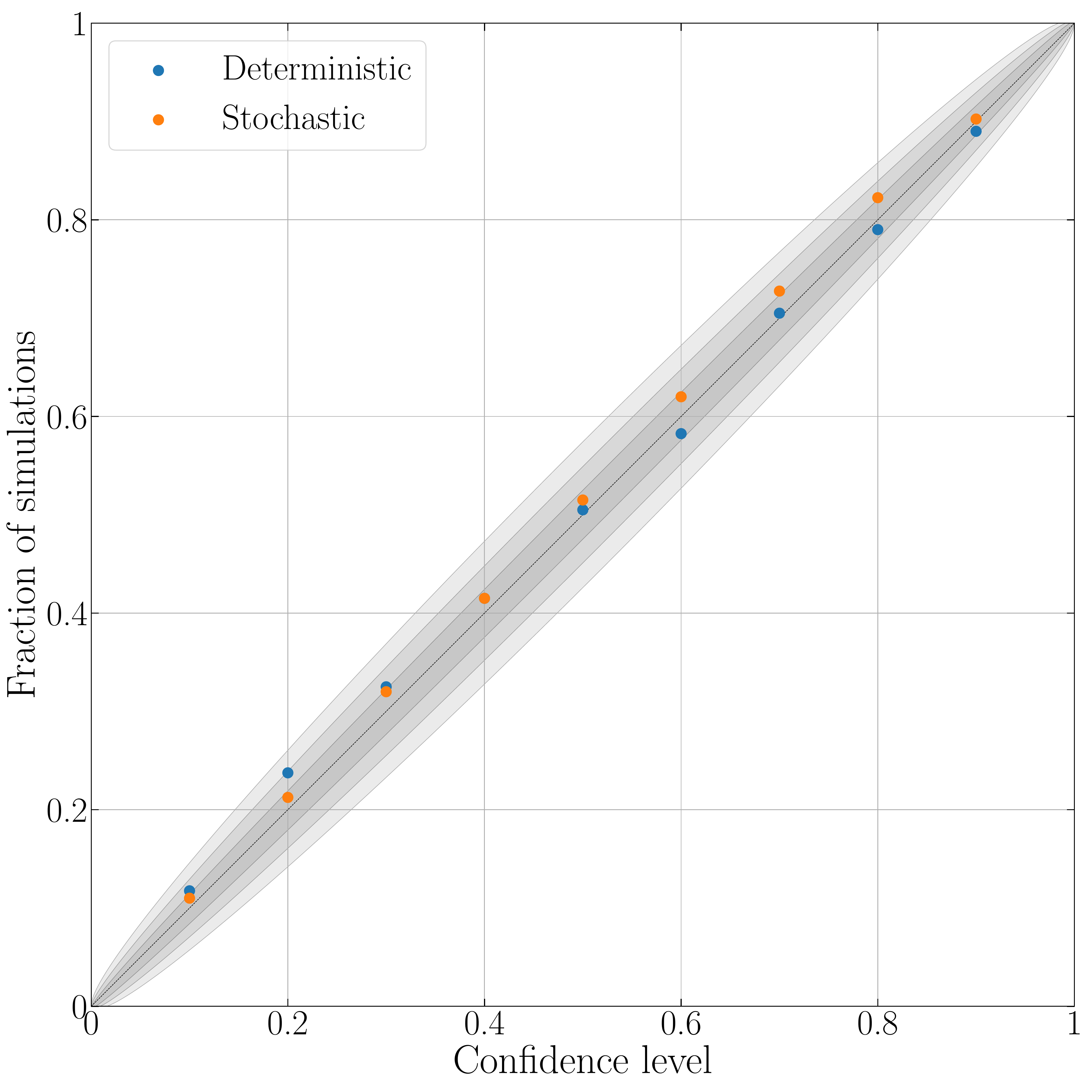}
        \includegraphics[width=0.8\columnwidth]{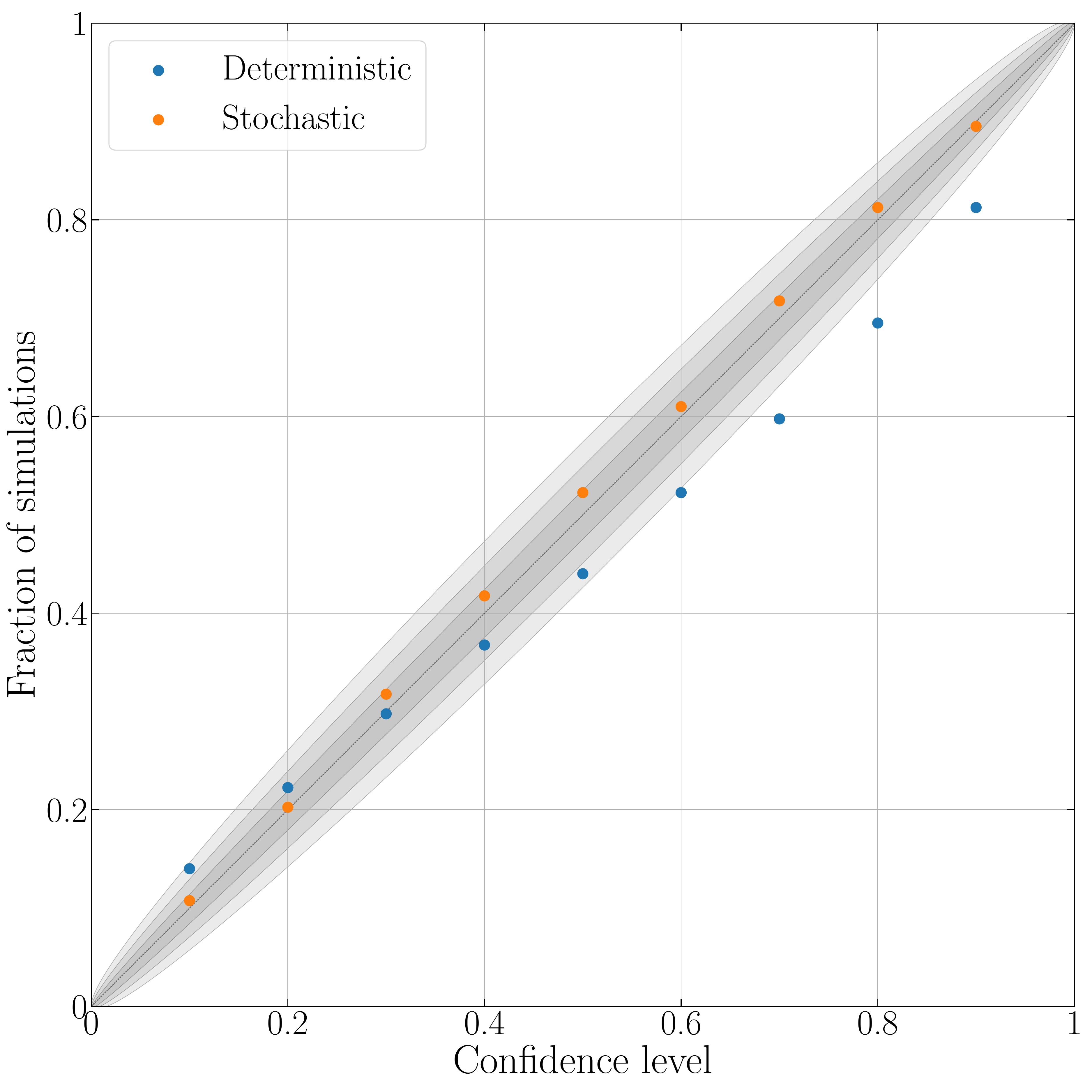}
        \includegraphics[width=0.8\columnwidth]{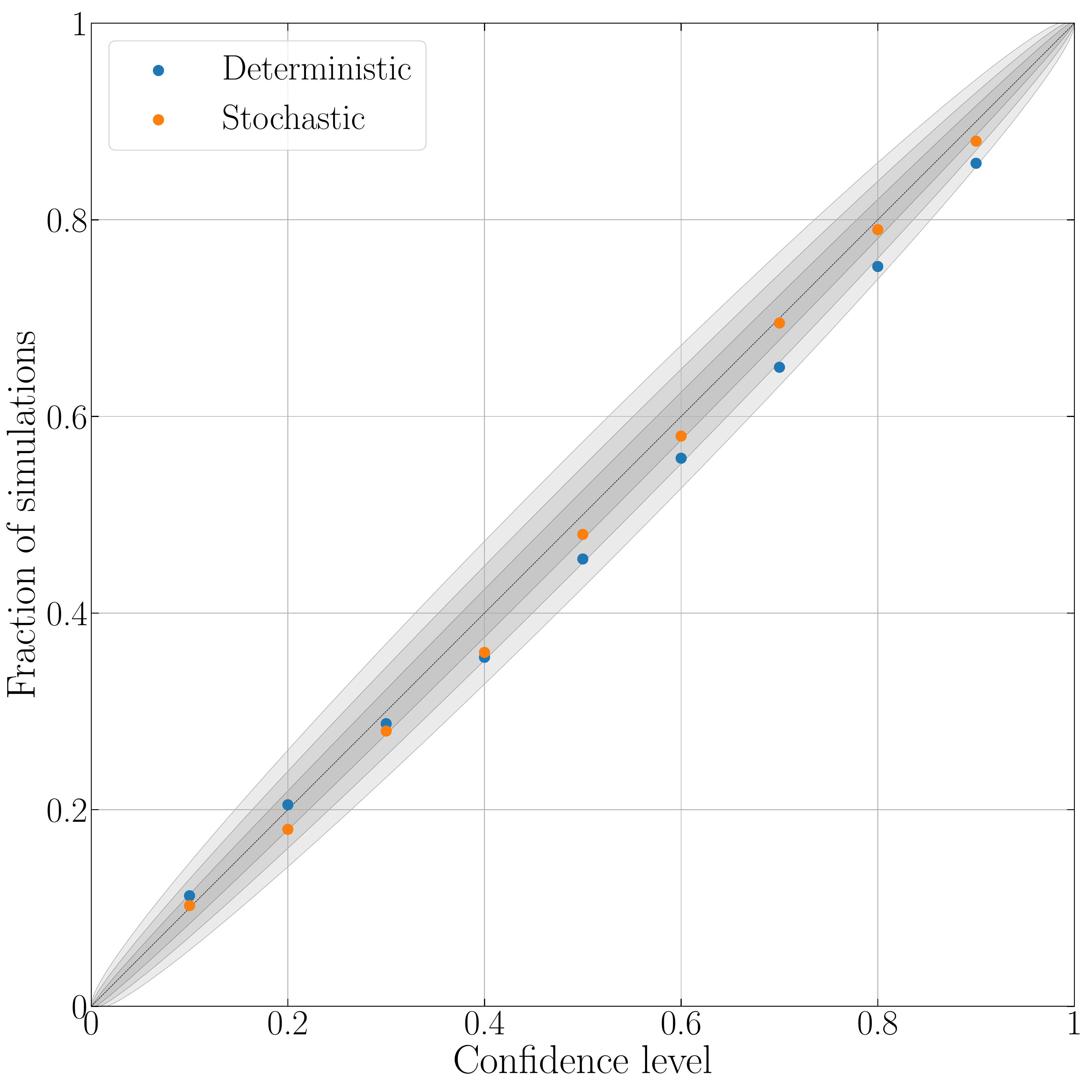}
        \caption{The fraction of simulations for which the upper bounds on $\epsilon_B$ calculated by the deterministic (blue) and stochastic (orange) models are above its true value for each confidence level among $0.1,~0.2,\dots,~0.9$. The upper left panel is for $(f_\mathrm{DM}, T, \epsilon_B) = (20\,\mathrm{Hz},~3.6\times10^3\,\mathrm{s},~1.1\times10^{-22})$, the upper right for $(20\,\mathrm{Hz},~1.8\times10^5\,\mathrm{s},~2.3\times10^{-23})$, the lower left for $(100\,\mathrm{Hz},~7.1\times10^2\,\mathrm{s},~5.1\times10^{-23})$, and the lower right for $(100\,\mathrm{Hz},~3.6\times10^4\,\mathrm{s},~1.1\times10^{-23})$.}
        \label{fig_pp}
    \end{center}
\end{figure*}

The upper bounds on $\epsilon_B$ were calculated for confidence levels of $0.1, 0.2, \dots, 0.9$ in the deterministic and stochastic models.
Figure~\ref{fig_pp} shows the fraction of simulations where the true value of $\epsilon_B$ is lower than the obtained upper bounds for each confidence level and each pair of $(f_\mathrm{DM}, T, \epsilon_B)$.
The gray regions represent the $1$--$\sigma$, $2$--$\sigma$, and $3$--$\sigma$ confidence intervals of statistical errors due to the finite number of simulations.
For $T = 5 \tau$, both models are within the confidence intervals.
For $T = \tau / 10$, the deterministic model shows statistically significant deviations from the diagonal line, while the stochastic model is well within the confidence intervals.
Thus, we need to include the stochastic effect to evaluate the upper limit of the coupling constant.

\section{Application for future upper bound}
\label{sec_experiment}

In this section, we apply our method to derive expected constraints by a future experiment.
Here we use an alternative to the observational data $\rho_{\mathrm{obs}}$ in order to give a rough estimate of the upper bound put by future experiments.
We numerically evaluate the future upper bound on velocity-dependent and independent signals.
Then, we consider an aLIGO-like experiment in the following analysis to investigate the stochastic effect on axion and dark photon DM signals.
Assuming the measurement time $T$, we use discretized frequencies $f_n = f_\mathrm{DM} +T^{-1}(n-1/2)$ in the following calculation.

\subsection{Future upper bound by frequentist's method}

\begin{figure}[tb]
  \centering
  \includegraphics[width=1.\columnwidth]{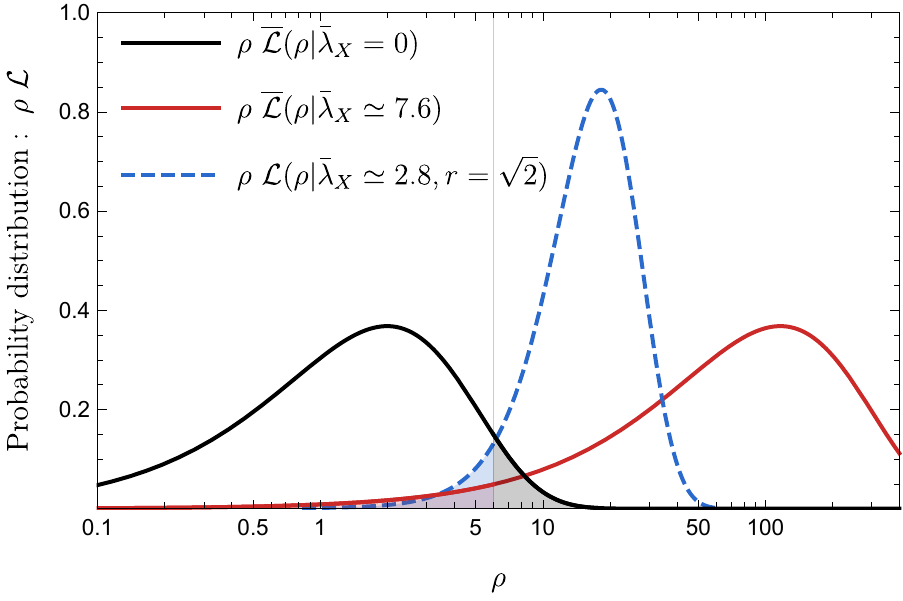}
  \caption{
  {
  Likelihood functions for single measurement $N_\mathrm{bin}=1$ with $\Delta_X=\Delta_s$ multiplied by $\rho$ for illustration purpose.
  We present the noise-only likelihood (black solid), the one marginalized over the DM stochasticity $r$ (red solid)
  and the reference case with a fixed $r=\sqrt{2}$ without maginalization (blue dashed), based on Eqs.~\eqref{eq_def_L} and \eqref{eq_barL_single}.
  The gray vertical line denotes $\rho_{\rm dt}\simeq 6.0$ above which the integral of the noise-only likelihood becomes $\alpha = 0.05$ (gray shaded region).
  The upper bounds on $\bar \lambda_X$ for the other two likelihoods are obtained such that their integrals for $\rho<\rho_{\rm dt}$ (red and blue shaded regions) are equal to $1-\beta = 0.05$ according to Eq.~\eqref{eq_def_lambdaup} with the replacement of $\rho_\mathrm{obs}$ by $\rho_\mathrm{dt}$. Since the marginalized likelihood has a more extended distribution for a higher $\rho$, we obtain a more conservative upper bound than the reference case. }
  }
  \label{fig_probdist_explain}
\end{figure}

 We assume that no detectable DM signal is included in an experimental data and the data is mostly contaminated with noise.
Although the typical size of the experimental noise is estimated by $S_\mathrm{noise}$, the observed data can accidentally become much larger than it. 
Considering this point, we introduce the false alarm rate $\alpha$ and the detection threshold $\rho_{\rm dt}$ assuming the background only case, $\bar\lambda_X = 0$, as
\begin{align}
	 \alpha 
	&=\int_{\rho_{\rm dt}}^\infty {\rm d} \rho ~\overline{\mathcal L}(\rho|\{0 \})
	=  \frac{\Gamma(N_\mathrm{bin},\rho_\mathrm{dt}/{2})}{\Gamma(N_\mathrm{bin})}
	,
	\label{eq_def_rhodt}
\end{align}
where  $\Gamma$'s are (incomplete) gamma function. 
$\alpha$ is the probability that the noise-only data exceeds the threshold value $\rho_\mathrm{dt}$. By inversely solving Eq.~\eqref{eq_def_rhodt}, $\rho_\mathrm{dt}$ for a given $\alpha$ can be determined. For a smaller false alarm rate $\alpha$, one obtains a higher threshold $\rho_\mathrm{dt}$.
In this paper, we adopt $\rho_\mathrm{dt}$ with $\alpha = 0.05$ as an alternative to $\rho_\mathrm{obs}$. 
Note that the concrete value of $\alpha$ is rather arbitrary and 0.05 is just an example.
The black line in Fig.~\ref{fig_probdist_explain} presents the noise-only likelihood function with $N_\mathrm{bin}=1$. 
The integration in Eq.~\eqref{eq_def_rhodt} corresponds to the gray shaded region, and the gray vertical line indicates $\rho_{\rm dt}\simeq 6.0$ for $\alpha = 0.05$ there.

Now we can compute the upper bound $\bar \lambda_{\rm up}$ in Eq.~\eqref{eq_def_lambdaup} by replacing $\rho_\mathrm{obs}$ by $\rho_\mathrm{dt}$ obtained above. 
For $N_\mathrm{bin}=1$ and $\Delta=\Delta_s$, we obtain $\bar \lambda_{\rm up}\simeq 7.6$ in a stochastic case. 
We also compute the upper bound in the deterministic case, where the DM stochasticity is not marginalized but fixed as $r=\sqrt{2}$, and we find a tighter constraint, $\bar \lambda_{\rm up}\simeq 2.8$.
As shown in Fig.~\ref{fig_probdist_explain}, this difference comes from the distribution of the likelihood functions.
The stochasticity of the DM amplitude broadens the likelihood function when it is taken into account by marginalization.
Then, for the same integrated value over the tail of its distribution, the marginalized one (red solid) is more shifted than the reference one (blue dashed), which implies a looser bound on the signal size $\bar \lambda_X$.

The translation from $\bar\lambda_\mathrm{up}$ into the upper bound on the DM coupling constant is straightforward by Eq.~\eqref{eq_def_barlambda}.
Using these formulas, we can determine the upper bound of the coupling constant including the noise and the stochastic effect, which is numerically discussed in  Sec.~\ref{sec_velocity_independent} and \ref{sec_velocity_dependent}.

Before we perform the numerical calculation, we show the analytic formula for $\bar\lambda_\mathrm{up} (T)$ derived in App.~\ref{sec_approx_measurmetn}:
\begin{align}
    &\bar\lambda_\mathrm{up} (T)
    =\nonumber
    \\&
    \begin{cases}
        (\Delta_{X,\mathrm{tot}})^{-1/2}
          \sqrt{\frac{\ln(\alpha)}{\ln(\beta)}-1} 
         ,
        &\mathrm{for}~ T<\tau/\kappa,
        \\
        (\Delta_{X,\mathrm{tot}})^{-1/2}
        \sqrt{M_\alpha+ M_{1-\beta}} \left(\kappa T/\tau \right)^{1/4}
        &\mathrm{for}~ T\gg\tau.
    \end{cases}
    \label{eq_lambdas_twolimit}
\end{align}
where $\Delta_{X,\mathrm{tot}}$ represents the sum of all $\Delta (f_n)$ over the whole frequency range $n$ as defined in Eqs.~\eqref{eq_delta_N_tot} and \eqref{eq_delta_pp_tot},
and their values are given by 
$\Delta_{s,\mathrm{tot}} = 1$, 
$\Delta_{\perp,\mathrm{tot}} \simeq 0.19$, and $\Delta_{\parallel,\mathrm{tot}} \simeq 0.62$.
$M_\chi$ represents the relation between the peak width and the area of the Gaussian distribution defined by 
\begin{align}
    \chi  = 
    \int_{M_\chi}^\infty \df z 
    \frac{1}{\sqrt{2\pi}} \exp(-z^2/2).
\end{align}
The analytic formulae correctly reflect the dependence on the measurement time as $\bar \lambda_\mathrm{up}\propto T^0$ for $T<\tau/\kappa$ and  
$\bar \lambda_\mathrm{up}\propto T^{1/4}$ for $T\gg \tau$.
We need the numerical calculation for the marginal region ($T\sim \tau$), which is discussed in the following section.  

Note that the derived upper bound  is more conservative  than one simply estimated by the signal-to-noise ratio (SNR) equal to one.
Indeed, SNR$=1$ corresponds to the $\bar\lambda_\mathrm{up}  = 1$ for $N_\mathrm{bin}=1$ while $\bar\lambda_\mathrm{up} (T)$ is usually larger than one for small $\alpha$ and $1-\beta$ in our estimate.
The difference comes from the stochastic effect and the conservative choice of  $\rho_\mathrm{obs}$, which leads to the looser upper bound.
When we conduct the measurement and compute the upper bound in Eq.~\eqref{eq_def_lambdaup} without the replacement
by $\rho_\mathrm{dt}$, the {actual} upper bound can be tighter.

\subsection{Velocity-independent signal}
\label{sec_velocity_independent}

Here, we numerically estimate the upper bound~\eqref{eq_def_lambdaup} for a velocity-independent signal, which is denoted by $\bar \lambda_{\mathrm{up}}^{(s)}(T)$ and a superscript `s' represents the velocity-independent signal.
The deterministic part of the spectral shape of velocity-independent signals is represented by $\Delta(f_n)=\Delta_s(f_n)$ in Eq.~\eqref{eq_def_Delta_N}, which reflects the velocity distribution of DM.

\begin{figure}[t]
  \centering
  \includegraphics[width=1.\columnwidth]{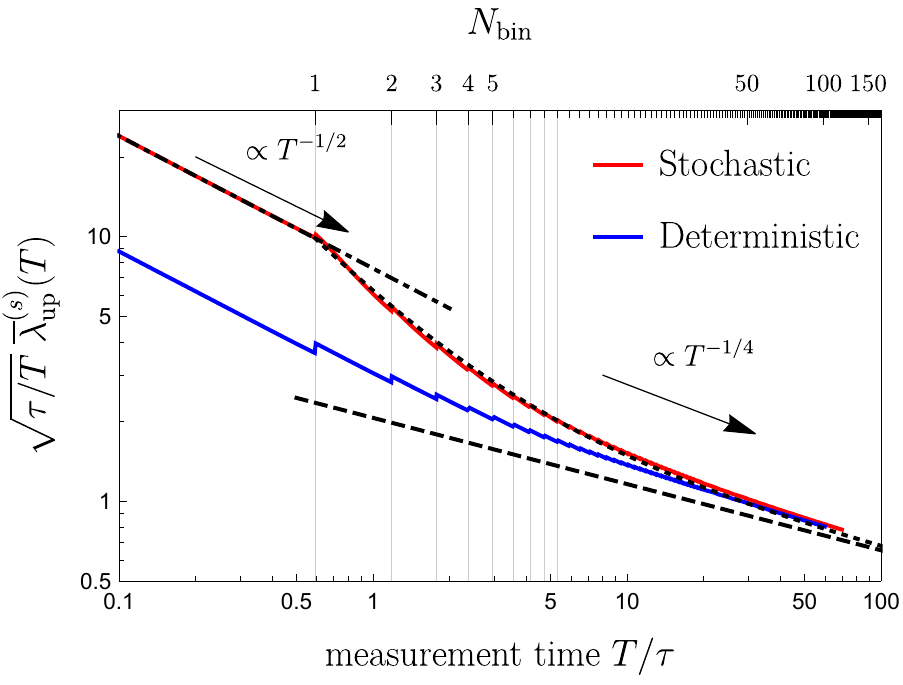}
  \caption{
  Upper bound on the coupling constant for velocity-independent signals.
  We perform the frequentist's method to estimate $\bar \lambda_{\mathrm{up}}^{(s)}(T)$ (red line) for the likelihood Eq.~\eqref{eq_label_likelihood} with $\Delta_X(f_n)=\Delta_s(f_n)$, $\alpha =1-\beta=0.05$ and $\kappa = 1.69$.
  The vertical axis describes the normalized upper bound of coupling constant.
  The blue line represents the upper bound for the deterministic case, where the fluctuation of the field amplitude is neglected and the likelihood is estimated by Eq.~\eqref{eq_likelihood_deterministic}.
  The dot-dashed and dashed lines represent the approximation formulas for short-time and long-time measurement Eq.~\eqref{eq_lambdas_twolimit}, respectively.
  $\bar \lambda_{\mathrm{up}}^{(s)}(T)$ is fitted by Eq.~\eqref{eq_lambdath_N}, which is shown as the dotted line.
  }
  \label{fig_lambdath_v_independent}
\end{figure}

By using the likelihood function, we perform the frequentist's method developed in Sec.~\ref{sec_likelihood} to compute $\bar \lambda_{\mathrm{up}}^{(s)}(T)$.
We show the upper bound of the coupling constant in Fig.~\ref{fig_lambdath_v_independent}, where the vertical axis describes $\bar\lambda_\mathrm{up} ^{(s)}(T)\sqrt{\tau/T}$ whose $T$ dependence is the same as $g_\mathrm{up}(T)$.
$\bar \lambda_{\mathrm{up}}^{(s)}(T)$ is calculated with $1-\beta=0.05$, $\rho_\mathrm{dt}$ for $\alpha =0.05$ and $\kappa = 1.69$. 
The approximate formulas in Eq.~\eqref{eq_lambdas_twolimit} are plotted as dot-dashed and dashed lines, where we use $\Delta_{s,\mathrm{tot}} \simeq 1$.
The upper bound improves by time, where $g_\mathrm{up}(T) \propto T^{-1/2}$ for $T\ll \tau$ and  $g_\mathrm{up}(T) \propto T^{-1/4}$ for $T\gg \tau$.
The numerical result (red line) smoothly connects two limiting cases. 
Since we treat the number of the frequency bins $N_\mathrm{bin}$ as an integer, $\bar\lambda_\mathrm{up} ^{(s)}$ has small jumps when $N_\mathrm{bin}$ changes.
Note that $N_\mathrm{bin}=1$~\eqref{eq_num_bin} is different from $T/\tau=1$  due to the choice of $\kappa\simeq 1.69$ in a velocity-dependent signal.

To compare the effect of the stochastic field value, we also show the upper bound calculated for the deterministic case (blue line), where the fluctuation of the field amplitude is neglected and the likelihood is estimated by Eq.~\eqref{eq_likelihood_deterministic}.
The stochastic and deterministic cases have the largest deviation for $N_\mathrm{bin}=1$, where the ratio of them is about 2.7, which is consistent with the previous result in Ref.~\cite{Centers:2019dyn}.
As the measurement time or the number of bins becomes larger, the difference between the stochastic and deterministic results shrinks.
It is because fluctuations of the field value becomes negligible compared to the instrumental noise when the coupling constant is severely constrained by a long measurement time.

For the later convenience, we  find the fitting formula of the numerically computed $\bar\lambda_\mathrm{up} ^{(s)}(T)$.
We require that $\bar\lambda_\mathrm{up} ^{(s)}(T)$ smoothly connects the asymptotic formulas in Eq.~\eqref{eq_lambdas_twolimit}.
We approximate the transient behavior by adding the power of $T$ in the following way:
 \begin{align}
     &\bar\lambda_\mathrm{up} ^{(s)}(T) =
     \nonumber\\&
     \begin{cases}
        \sqrt{\tfrac{\ln(\alpha)}{\ln(\beta)}-1}  
         &\mathrm{for}~T<\tau/\kappa,
         \\
         \sqrt{M_\alpha+ M_{1-\beta}} \left(    \kappa\frac{T}{\tau } \right)^{1/4}&
         \\+
         (\kappa T/\tau)^q \left[
         \sqrt{\tfrac{\ln(\alpha)}{\ln(\beta)}-1}   - 
         \sqrt{M_\alpha+ M_{1-\beta}} 
         \right]
         &\mathrm{for}~T>\tau/\kappa, 
     \end{cases}
     \label{eq_lambdath_N}
 \end{align}
 where the index $q$ is determined by the least squares method.
 The fitting value is $q\simeq -0.61$, and it fits well the numerical result (red line) as shown by a black dotted line in Fig.~\ref{fig_lambdath_v_independent}.

\subsection{Velocity-dependent signal}
\label{sec_velocity_dependent}

Next, we numerically estimate the upper bound~\eqref{eq_def_lambdaup} for the velocity-dependent signal, $\bar \lambda_{\mathrm{up}}^{(v)}(T)$. 
The main difference from the velocity-independent signal is the spectral shape.
The deterministic part of the spectral shape of velocity-dependent signals is represented by $\Delta_X(f_n)=\Delta_x(f_n)+\Delta_y(f_n)$, which depends on the relative direction of the interferometer arms to the solar velocity.
Here, we consider the two typical directions introduced below Eq.~\eqref{eq_sspace}:
\begin{align}
    \Delta_X=
    \begin{cases}
        2 \Delta_\perp
         &\mathrm{for~conservative~direction},
         \\
        \Delta_\parallel+\Delta_\perp
         &\mathrm{for~optimal~direction}.
    \end{cases}
    \label{eq_Delta_concervative_optimal}
\end{align}
Their integrated signals are given by 
\begin{align}
    &\Delta_{X,\mathrm{tot}}=
    \nonumber\\
    &\begin{cases}
         2\Delta_{\perp,\mathrm{tot}}\simeq 0.38
         &\mathrm{for~conservative~direction},
         \\
        \Delta_{\perp,\mathrm{tot}}+ \Delta_{\parallel,\mathrm{tot}}\simeq 0.81
         &\mathrm{for~optimal~direction}.
    \end{cases}
\end{align}

\begin{figure}[t]
  \centering
  \includegraphics[width=1.\columnwidth]{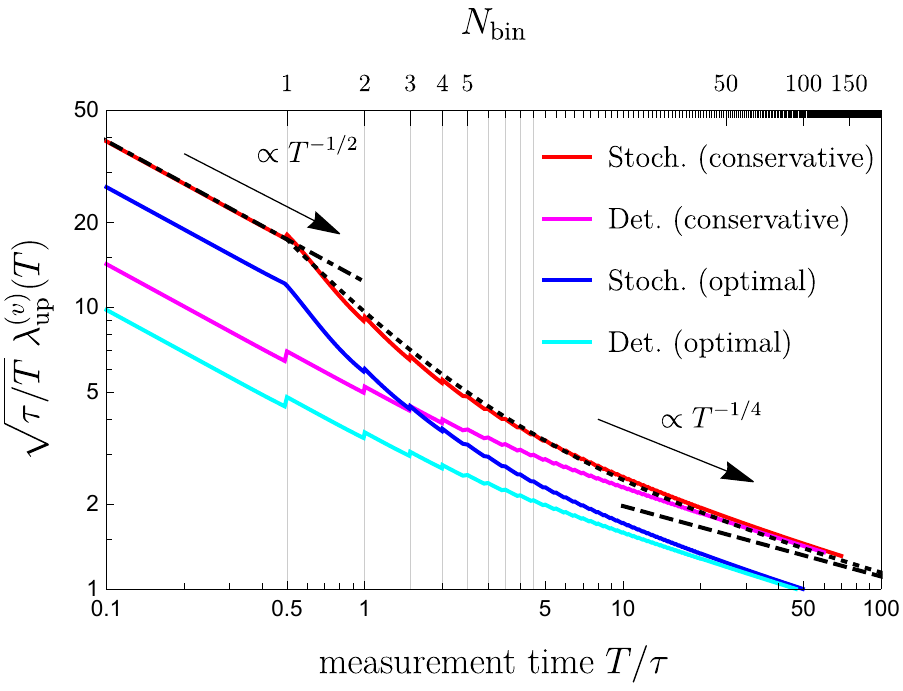}
  \caption{
  The same plot as Fig.~\ref{fig_lambdath_v_independent} for the velocity-dependent signal.
  We perform the frequentist's method to estimate $\lambda_{\mathrm{up}}^{(v)}(T)$ (red and blue lines) for the likelihood Eq.~\eqref{eq_label_likelihood} with $\alpha =1-\beta=0.05$, $\kappa = 2$, and $\Delta(f_n)$ defined in Eqs.~\eqref{eq_Delta_concervative_optimal}.
  The red and blue solid lines represent the upper bound 
  when the Sun moves relative to the interferometer arms in a conservative and an optimal directions, respectively.
  The magenta and cyan lines represent the upper bounds for the deterministic case, where the fluctuation of the field amplitude is neglected and the likelihood is estimated by Eq.~\eqref{eq_likelihood_deterministic}.
  The dot-dashed and dashed black lines represent the approximation formulas for short and long measurement time in Eq.~\eqref{eq_lambdas_twolimit}, respectively.
  $\lambda_{\mathrm{up}}^{(v)}(T)$ is fitted by Eq.~\eqref{eq_lambda_v_fitting}, which is shown by the dotted black line.
  }
  \label{fig_lambdath_v_dependent}
\end{figure}

Again, we perform the frequentist's method in Sec.~\ref{sec_likelihood} to estimate $\bar \lambda_{\mathrm{up}}^{(v)}(T)$ with $\alpha =1-\beta=0.05$ and $\kappa = 2$.
The results are shown in  Fig.~\ref{fig_lambdath_v_dependent}.
The two upper limits (red and blue lines) are estimated based on these two directions.
The optimal direction leads to about $\sqrt{0.81/0.38}\simeq 1.5$ times stringent upper bound than that of the conservative  direction.
However, the interferometer arms can take the optimal direction only in limited instances, if any, due to the rotation of the Earth.
To be conservative, we focus on the upper bound for the conservative direction in the following calculation.

In the same way to the velocity-independent signal, we compare the stochastic case to the deterministic case (magenta and cyan lines) estimated by Eq.~\eqref{eq_likelihood_deterministic}.
Since the probability distribution is the Rayleigh distribution both for the velocity-independent and velocity-dependent signals, the stochastic effect for the  velocity-dependent signal is similar to that for the velocity-independent signals in Fig.~\ref{fig_lambdath_v_independent}.

For later convenience, we estimate the fitting formula of $\bar\lambda_\mathrm{up} ^{(v)}(T)$ in the same way as the case of the velocity-independent signal.
The fitting formula for the velocity-dependent signal with the conservative direction is given by
 \begin{align}
     &\bar\lambda_\mathrm{up} ^{(v)}(T) \simeq
     \nonumber\\&
     \begin{cases}
        (\Delta_{{X,\mathrm{tot}}})^{-1/2}
        \sqrt{\tfrac{\ln(\alpha)}{\ln(\beta)}-1}  
         &\mathrm{for}~T<\tau/\kappa,
         \\
          (\Delta_{{X,\mathrm{tot}} })^{-1/2}
          \big(
             \sqrt{M_\alpha+ M_{1-\beta}} \left(    \kappa\frac{T}{\tau } \right)^{1/4}
             &
    \\
             +
             (\kappa \tfrac{T}{\tau})^q \left[
             \sqrt{\tfrac{\ln(\alpha)}{\ln(\beta)}-1}   - 
             \sqrt{M_\alpha+ M_{1-\beta}} 
             \right]
         \big)
         &\mathrm{for}~T>\tau/\kappa, 
     \end{cases}
     \label{eq_lambda_v_fitting}
 \end{align}
 with $\kappa=2$, $\Delta_{{X,\mathrm{tot}}} \simeq 0.38$, and $q\simeq -0.61$.
 The fitting formula is shown as a black dotted line in Fig.~\ref{fig_lambdath_v_dependent}.

\subsection{Constraint on axion DM}
\label{sec_sensitivity_axion}

In the aLIGO-like experiment, the axion signal appears in two detection ports, a transmission and reflection port, where the former is more sensitive to the low-mass range since the photons travel cavity for an odd-number of times~\cite{Nagano:2021kwx}.
Using Eqs.~\eqref{eq_def_barlambda} and \eqref{eq_lambdath_N}, 
we translate $\bar \lambda_\mathrm{axion}< \bar\lambda_\mathrm{up} ^{(s)}$ into the upper bound of coupling constant as 
\begin{align}
    &g_a < 
    g_{a,\mathrm{up}}(T)
    \equiv
     \frac{ 8\sqrt{2} \pi }{ m \sigma_a(2\pi/k) }
    \frac{ \sqrt{T S_{\rm noise}(f_\mathrm{DM})}}{2  T } 
    \bar\lambda_\mathrm{up} ^{(s)}(T)
   \nonumber \\
    &\sim 
    5\times 
    10^{-11} \mathrm{GeV}^{-1}
    ~
    \bar\lambda_\mathrm{up} ^{(s)}(T)
    \frac{1064~\mathrm{nm}}{(2\pi/k)}
    \nonumber\\&\quad\times 
    \left(
    \frac{0.4~\mathrm{GeV}/\mathrm{cm}^3}{\rho_\mathrm{DM}}
    \frac{S_{\rm noise}(f_\mathrm{DM})}{ 10^{-40} ~\mathrm{Hz}^{-1}  }
    \frac{1~\mathrm{day}}{T}
    \right)^{1/2}
     .
     \label{eq_upper_coupling_axion}
\end{align}

\begin{figure}[t]
  \centering
  \includegraphics[width=1.0\columnwidth]{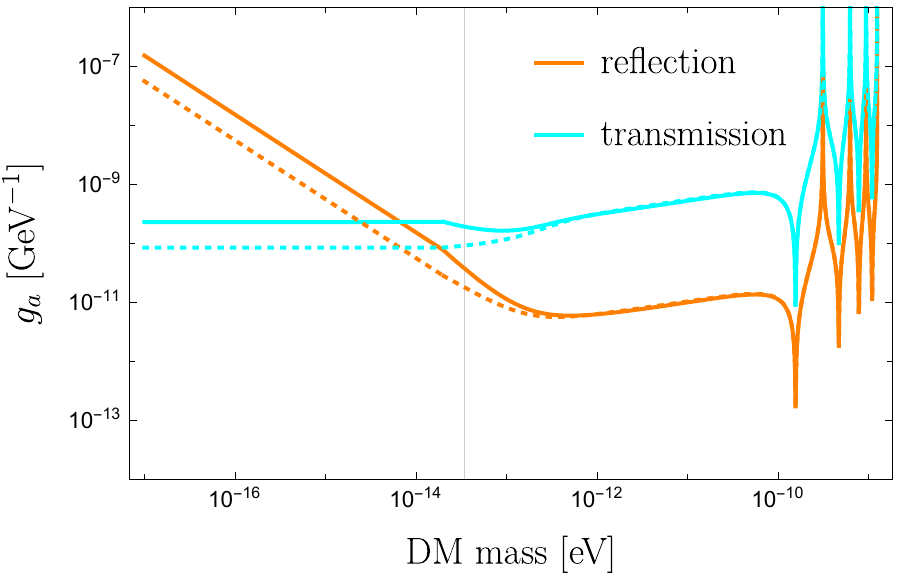}
  \caption{
  Upper bound of coupling constant $g_a$ for axion DM with 95\%CL ($\alpha = 1-\beta = 0.05$).
  We assume the aLIGO-like detector with a measurement time $T=$1 day, where the response function is calculated in Ref.~\cite{Nagano:2021kwx}. 
  The solid and dotted lines represent the future exclusion limit with and without the stochastic effect of the field value, respectively.  
  The orange and cyan lines represent the sensitivity by the transmission and reflection ports.
  The gray vertical line represents a mass at $T = \tau$.
  }
  \label{fig_Axion_aLIGO_const_1dayy}
\end{figure}

We show the future constraint on $g_a$ for an aLIGO-like experiments with a laser wavelength $2\pi/k = 1064$~nm and $T=1$~day in Fig.~\ref{fig_Axion_aLIGO_const_1dayy}, where the noise spectrum is calculated in Ref.~\cite{Nagano:2021kwx}.
The orange and cyan lines represent the sensitivity by the transmission and reflection ports, respectively.
The solid and dotted lines represent the upper bound by the stochastic and deterministic cases.
The upper bound becomes looser for the smaller axion mass than $10^{-15}$~eV due to the stochastic effect.

Compared to the previous results~\cite{Nagano:2021kwx}, our results put the looser constraint for the following two reasons.
At first, we estimate the future exclusion limit by frequentist's method while the Ref.~\cite{Nagano:2021kwx} estimates the future sensitivity with $\mathrm{SNR}=1$, which leads a difference of about factor 2.
Second, as we mentioned, the stochastic fluctuation of field value loosen the upper bound by about factor 3, which is shown by the difference between the solid and dotted lines.
Due to these reasons, we estimate the conservative upper bound for the aLIGO-like experiment.

\subsection{Constraint on dark photon DM}
\label{sec_sensitivity_dark_photon}

The interferometer experiment on dark photon DM has three different signals, $s_\mathrm{time}$,  $s_\mathrm{space}$, and  $s_\mathrm{charge}$.
The upper bound of the coupling constant is placed by the strongest one among them:
\begin{align}
    \epsilon_D < \min\left(
    \epsilon_{D,\mathrm{up}}^\mathrm{(time)} ,
    \epsilon_{D,\mathrm{up}}^\mathrm{(space)},
    \epsilon_{D,\mathrm{up}}^\mathrm{(charge)}
    \right).
\end{align}

As we derived in Sec.~\ref{sec_vector}, $s_\mathrm{time}$ and $s_\mathrm{charge}$ are the velocity-independent signals and have similar power spectrum to the axion except for the constant coefficient. 
On the other hand, $s_\mathrm{space}$ is a velocity-dependent signal, and its spectral distribution is different from that of the axion.
Then, the upper bounds of coupling constant are given by
\begin{align}
    \epsilon_{D,\mathrm{up}}^\mathrm{(time)} 
    &=
    \left(
         e
        \frac{(Q/M)_\mathrm{in}   }{mL}
        \sqrt{\frac{2\rho_{\rm DM}}{3m^2}  }
        \sin^2\left(\frac{mL}{2}\right)
     \right)^{-1}
\nonumber\\&\times
     \frac{ \sqrt{T S_{\rm noise}(f_\mathrm{DM})}}{2  T } 
      \bar\lambda_\mathrm{up} ^{(s)}(T)
     ,
\\
    \epsilon_{D,\mathrm{up}}^\mathrm{(charge)} 
    &=
    \left(
        e
        \frac{|(Q/M)_e-(Q/M)_\mathrm{in}| }{2Lm}
        \sqrt{\frac{2\rho_{\rm DM}}{3m^2}  }
     \right)^{-1}
\nonumber\\&\times
     \frac{ \sqrt{T S_{\rm noise}(f_\mathrm{DM})}}{2  T } 
     \bar\lambda_\mathrm{up} ^{(s)}(T),
\\
    \epsilon_{D,\mathrm{up}}^\mathrm{(space)} 
    &=
    \left(
        e
        \frac{ (Q/M)_\mathrm{in}   \bar v}{2 {\sqrt{2}}} 
        \sqrt{\frac{2\rho_{\rm DM}}{3m^2}  }
     \right)^{-1}
\nonumber\\&\times
     \frac{ \sqrt{T S_{\rm noise}(f_\mathrm{DM})}}{2  T } 
      \bar\lambda_\mathrm{up} ^{(v)}(T)
     ,
\end{align}
where $\bar\lambda_\mathrm{up} ^{(v)}(T)$ is evaluated by Eq.~\eqref{eq_lambda_v_fitting}.

\begin{figure}[t]
  \centering
  \includegraphics[width=1.0\columnwidth]{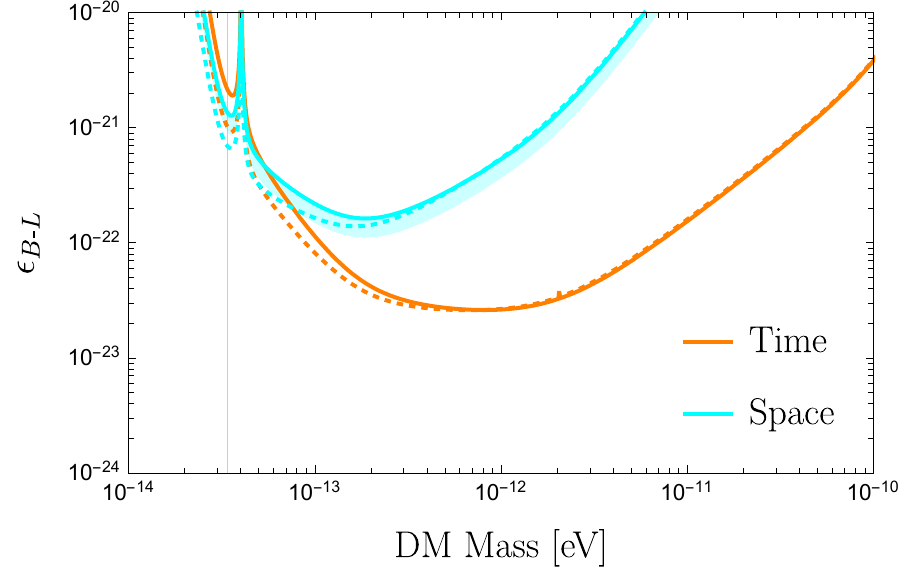}
  \caption{
  The upper bound of coupling constant $\epsilon_{B\mathchar`-L}$ for dark photon DM with 95\%~CL ($\alpha = 1-\beta = 0.05$).
  We assume the aLIGO like detector with a measurement time $T=1$~day, where the noise spectrum is calculated in Ref.~\cite{aligosensitivity,Morisaki:2020gui}. 
   The solid and dotted lines represent the future exclusion limit with and without the stochastic effect of the field value, respectively.  
  The orange and the cyan lines represent the upper bound from the temporal and spatial modulation of mirrors induced by dark photon DM.
  The gray vertical line represents a mass at $T = \tau$.
  The cyan line describes the conservative direction of the solar velocity while the optimal direction put a severe constraint (cyan shaded region). 
  }
  \label{fig_DPhoton_aLIGO_const_1day}
\end{figure}

For the aLIGO-like experiment, the arm length is $L= 4\times 10^3$~m and mirrors made of the same ingredients, $q_\mathrm{in}=q_\mathrm{e}\simeq 0.5/m_n$ for $D=B$-$L$ charge.
In this case, the upper bound is put by both $\epsilon_{B\mathchar`-L,\mathrm{up}}^\mathrm{(time)}$ and $\epsilon_{B\mathchar`-L,\mathrm{up}}^\mathrm{(space)}$.
We estimate the upper bound by adopting the same noise spectrum as Ref.~\cite{Morisaki:2020gui} and $\bar\lambda_\mathrm{up} ^{(v)}$ for the conservative direction.
We show the future constraints on $\epsilon_{B\mathchar`-L}$ with $T=1$~day in Fig.~\ref{fig_DPhoton_aLIGO_const_1day}, where the orange and cyan lines represent $\epsilon_{B\mathchar`-L,\mathrm{up}}^\mathrm{(time)} $ and $\epsilon_{B\mathchar`-L,\mathrm{up}}^\mathrm{(space)} $, respectively.
{
Since we use the fitting formula for $\bar\lambda_\mathrm{up} ^{(v)}$, 
the small jumps of $\bar\lambda_\mathrm{up} ^{(v)}$ in Fig.~\ref{fig_signal_vector_compare} disappear in Fig.~\ref{fig_DPhoton_aLIGO_const_1day}.
}
As we show in Fig.~\ref{fig_signal_vector_compare}, $\epsilon_{B\mathchar`-L,\mathrm{up}}^\mathrm{(space)} $ dominates the upper bound for small-mass region.
{
The dotted lines describes the upper bounds without the stochastic effect; it converges to the solid line for large mass since the number of bins increases as  $N_\mathrm{bin}\propto m$.
Although Fig.~\ref{fig_DPhoton_aLIGO_const_1day} describes a slight difference between solid and dotted lines for $m\gg 10^{-12}$~eV, the difference comes from a errors on the fitting function of $\bar\lambda_\mathrm{up} ^{(v)}$.
}

Compared to the previous work~\cite{Morisaki:2020gui}, our estimate predicts looser constraints due to some different calculations.
First, Ref.~\cite{Morisaki:2020gui} estimate the sensitivity by $\mathrm{SNR}=1$ while we calculate the future exclusion limit by frequentist's method.
Second, the stochastic effect loosens our constraints.

Ref.~\cite{LIGOScientific:2021odm} investigated the constraints on dark photon DM with almost one year data of LIGO and Virgo between $10-2000$~Hz. 
Let us revisit their results considering our discussion.
First, the coherent time of these DMs is much shorter than the observation time, e.g. the coherent time is shorter than one day for DM with a mass corresponding to $10$~Hz oscillation.
Thus, the stochastic effect of amplitude hardly affects their results.
Second, Ref.~\cite{LIGOScientific:2021odm}  assumed the isotropic velocity distribution of the dark photon dark matter, and they estimated the dark photon signal from the spatial gradient [Eq.~\eqref{eq_sspace}] by integrating over the all propagating direction and polarization directions.
Actually, the peculiar velocity of the Sun might affect the signal as we discussed in Sec.~\ref{sec_velocity_dependent}.
The direction dependence, however, is averaged by the rotation of the Earth.
A detailed analysis including a rotation of the Earth is left in a future work.

We comment on the daily modulation of signals.
Although we assume that $\vec v_\odot$ are constant in the above discussion, $\vec v_\odot$ and $\vec A$ actually change due to the rotation and revolution of the Earth. 
When the measurement time is much smaller than one day, 
we can safely neglect its rotation.
If not, the daily modulation could modify the sensitivity and upper bound of an interferometer experiment. 
Since we put the upper bound by the most conservative direction, our results provide the reliable upper bound even with the daily modulation.
We leave the detailed treatment of the daily modulation to a future work.

\section{Conclusion}
\label{sec_conclusion}

The ultralight bosonic field is a fascinating candidate of DM and intensively searched by various experiments.
Understanding the characteristics of DM signals is crucial for these experiments to properly put the upper limit on the coupling constant.
In this paper, we investigated the stochastic nature of bosonic DM fields including the dark photon DM and evaluated the upper bound of a DM coupling constant.

The ultralight bosonic field is described by the superposition of classical waves, which results in the stochastic amplitude and phase. 
In Sec.~\ref{sec_Stochastic_DM}, we estimated the probability distribution of the field value in frequency space.
We extended Ref.~\cite{Foster:2017hbq} to include the spatial derivative of the field value, which is closely related to the dark photon DM signal in interferometer searches.
We found that a power spectrum of the spatial derivative has a different spectral shape compared to that of the field value.
Using these probability distributions, we derived likelihood functions of axion and dark photon DM signals in interferometer searches.
We numerically simulated the dark photon DM signals to confirm  our semi-analytic calculations. 

Based on the frequentist's method, we can easily translate a power spectrum of interferometer searches to the upper bound of the coupling constant through Eqs.~\eqref{eq_label_likelihood} and ~\eqref{eq_def_lambdaup} including the stochastic nature of  DM signals. 
Next, we apply our formalism to estimate the upper bound by the future interferometer experiments.
We estimated the typical experimental data without DM signals from the projected experimental noise.
Then, we perform the frequentist's method to put the projected upper bound.
The normalized upper bounds are shown in Fig.~\ref{fig_lambdath_v_independent} and~\ref{fig_lambdath_v_dependent}. 
For the velocity-dependent and velocity-independent cases, 
the stochasticity on the amplitude loosens the upper limit up to about a factor 3, which is consistent with the previous work~\cite{Centers:2019dyn}.
Although Ref~\cite{Centers:2019dyn} focused on the shorter measurement time than the coherent time, we extend their analysis to the longer measurement time.
We found that as the measurement time exceeds the coherent time, the stochastic effect becomes inefficient.
We explicitly derived the time dependence of the upper limit, $\propto T^{-1/2}$ for a measurement time smaller than the coherent time and $\propto T^{-1/4}$ for a measurement time larger than the coherent time.

The expected constraints from aLIGO like experiments are shown in Fig.~\ref{fig_Axion_aLIGO_const_1dayy} for axion DM and Fig.~\ref{fig_DPhoton_aLIGO_const_1day} for dark photon DM.
Our analysis works for both axion and dark photon DM in a similar way.
The stochastic nature of the DM especially affects constraints on a small mass region due to long coherent time.
Our formulation can be applied to other experiments searching the ultralight bosonic DM including axion and dark photon DMs.
As future experiments will search low frequency  (small mass) regions, the stochastic effect becomes more important.

\section*{Acknowledgement}
\label{Acknowledgement}

We thank  Peter Wolf for meaningful comments.
In this work, HN, SM, TF, JK, YM, KN, and IO are supported by 
the JSPS KAKENHI Grant 
No.~JP17J01176, 
JP20J01928 (KN), 
JP20J21866 (JK)
JP19J21974 (HN),
and JP20H05859 (IO),
JSPS Grant-in-Aid for Scientific Research (B) No.~18H01224 (YM), 
Grant-in-Aid for Transformative Research Areas (A) No.~20H05850 and No.~20H05854 (YM),  
JST PRESTO Grant No.~JPMJPR200B (YM) and 
Grant-in-Aid for JSPS Research Fellow No.~17J09103 (TF),
Advanced Leading Graduate Course for Photon Science (HN),  
the Leading Graduate Course for Frontiers of Mathematical Sciences and Physics (JK),
NSF PHY-1912649 (SM), 
and JSPS Overseas Research Fellowship (IO),
respectively.

\appendix

\section{Fourier transformation of field value and its derivative}
\label{app_fourier_trans_field_values}

In this section, we derive the Fourier transformed field value~\eqref{eq_phi_fj_superposed} and its derivative~\eqref{eq_def_Del_Phi}.
We formally perform the summation in Eq.~\eqref{eq_phi_superposed}. 
At first, we reorganize the summation over the label $i$ in Eq.~\eqref{eq_phi_superposed} by
introducing three labels $(n,l,p)$ as
\begin{align}
     \Phi(t,\vec x)
    &=
    \sigma_\phi N_\phi^{-1/2}
    \sum_n
    \sum_l
    \sum_p^{\Delta N_{n,l}}
    \nonumber\\&\quad\times
    \cos\left(
    m(1+v_n^2/2)t 
    +m v_n \vec e_l\cdot \vec x
    +\theta_p^{(n,l)}
    \right),
    \label{eq_phi_sum_before}
\end{align}
where $n$ is the label of DM speed $|\vec v|=v_n$, $l$ is the label of the direction of DM velocity $\vec v = v_n \vec e_l$, and 
$p$ labels the oscillation phase of the contributions with the same $n$ and $l$.
Each label $(n,l)$ represents partial waves within a finite-size bin of the velocity $\Delta v_n \Delta \Omega_l$, where $\Delta v_n$ and $\Delta \Omega_l$ represent ranges of the speed and direction of partial waves.
Note that the $\Delta v_n$ is related to the frequency resolution of an actual experiment as 
$\Delta v_n = v^{-1}\Delta f_n/f$ 
with $2\pi f_n = m(1+v_n^2/2)$ in a non-relativistic limit, $v_n\ll 1$.
$\Delta N_{n,l}$ represents the number of waves that belongs to the $(n,l)$-th bin.
Using the velocity distribution in  Eq.~\eqref{eq_SHM_vec},  $\Delta N_{n,l}$ is given by
\begin{align}
    \Delta N_{n,l}
    &= N_\phi
    \int_{\Delta v_n}\df v 
     v^2
    \int_{\Delta \Omega_l} \df^2\Omega_e
    ~
    f_{\rm SHM}(\vec v(f,\vec e) + \vec v_\odot)
    .\label{eq_DNnl}
\end{align}
The summation over the index $p$ in Eq.~\eqref{eq_phi_sum_before} is computed as 
\begin{align}
    &\sum_p^{\Delta N_{n,l}}
    \cos\left(
    m(1+v_n^2/2)t 
    +m v_n \vec e_l\cdot \vec x
    +\theta_p^{(n,l)}
    \right)
\nonumber \\& =
    \Re \bigg[
    \exp(
    im(1+v_n^2/2)t 
    +im v_n \vec e_l\cdot \vec x
    )
\nonumber \\& \quad\times
    \sum_p^{\Delta N_{n,l}}
   \exp\left(i\theta_p^{(n,l)}\right)
   \bigg].
\end{align}
The summation of random phases can be understood as the random walk in two dimensional space, which is described by Gaussian variables in the limit of large $\Delta N_{n,l}$ as
$
     \sum_j^{\Delta N_{n,l}}
   \exp(i\theta_j)
   = 
   \sqrt{\Delta N_{n,l}/2}
   \left(
   R_{n,l} +i I_{n,l}
    \right)
    ,
$
where $R_{n,l}$ and $I_{n,l}$ follow the standard Gaussian distribution.
We can further rewrite them in the polar coordinate as 
$
r_{n,l} e^{i\theta_{n,l}}
\equiv
R_{n,l} +i I_{n,l}
$
, where the phase $\theta_{n,l}$ follows a uniform distribution over $[0,2\pi]$, and the radius $r_{n,l}$ follows the standard Rayleigh distribution~\eqref{eq_Rayleigh}.
Using these stochastic variables, the total field value is written as~\cite{Foster:2017hbq,Centers:2019dyn}
\begin{align}
     \Phi(t,\vec x)
    &=
    \sum_n \Phi_n(t,\vec x),
\nonumber\\
    \Phi_n(t,\vec x)
    &=
    \sigma_\phi 
    ~
    \sum_l
    \sqrt{
    \frac{\Delta N_{n,l}}{N_\phi}
    }
    ~
    \nonumber\\&\quad\times
    \frac{r_{n,l}}{\sqrt{2}}
    \cos\left(m(1+v_n^2/2)t +m v_n \vec e_l\cdot \vec x
    +\theta_{n,l} \right)
    \label{eq_phi_after_sum_i}
    .
\end{align}
An experiment searching for an ultralight scalar field often analyzes the Fourier-transformed data to extract the periodic signal.
Provided that we have the time-series data of the field value over the observation time $T$ spanning $t=[-T/2,T/2]$,
we find the Fourier-transformed field value at the $n$-th frequency bin $\tilde{\Phi}_n$ as
\begin{align}
    &\tilde \Phi_n(f,\vec x)
    \equiv
    \int^{T/2}_{-T/2} \df t\, e^{-2\pi i f t}
    \Phi_n(t,\vec x)
    \nonumber\\&
   \simeq
    \begin{cases}
        \dfrac{T\sigma_\phi}{2}
       \sum_l
        \sqrt{
        \dfrac{\Delta N_{n,l}}{N_\phi}
        }
        &
        \dfrac{r_{n,l}}{\sqrt{2}}
        \exp\left(
        im \vec v(f,\vec e_l)\cdot \vec x
        +i\theta_{n,l}
        \right)
        \\
        &
     (|f- f_n| \lesssim 1/T) \\
        0 \qquad\qquad&
        (|f- f_n| \gg 1/T)
    \end{cases}
    .
\end{align}
with $\vec v(f,\vec e) \equiv \sqrt{2(2\pi f/m -1)} \vec e$.
Here we approximated the time integral by the leading contribution at $f=f_n$,
while $\tilde \Phi_n(f,\vec x)$ is extended in the Fourier space over the frequency resolution $\Delta f = 1/T$.
Note that since each $\tilde{\Phi}_n$ contributes to the total field value $\tilde{\Phi}\equiv \sum_n \tilde{\Phi}_n$ at different frequency $f_n$, $\tilde{\Phi}(f_n)$ is equal to $\tilde{\Phi}_n(f_n)$.
Therefore, we obtain a master formula for the stochastic bosonic field as
\begin{align}
    &\tilde \Phi(f_n,\vec x) 
    \simeq 
    \nonumber\\&
    \frac{T}{2}  \sigma_\phi  
    ~ \sum_l   
    \sqrt{\frac{\Delta N_{n,l}}{N_\phi}}
    ~ 
    \left[
    \frac{r_{n,l}}{\sqrt{2}}
    \exp\left( im \vec v(f_n,\vec e_l)\cdot \vec x +i\theta_{n,l}\right)
    \right]
    ,
\label{eq_master_fieldvalue}
\end{align}
where the number of waves with the same velocity vector $v_n \vec e_l$ can be computed as
\begin{align}
     &\Delta N_{n,l}
    \nonumber
    \\&= N_\phi
    \int_{f_n-\Delta f/2}^{f_n+\Delta f/2} 
     v^2
     \frac{\df v}{\df f} \df f
    \int_{\Delta \Omega_l} \df^2\Omega_e
    ~
    f_{\rm SHM}(\vec v(f,\vec e) +  \vec v_\odot)
    .
\end{align}
This result is an extension of the formula in Ref.~\cite{Foster:2017hbq}, because it also describes the velocity direction, which is important for the signal induced from the spatial derivative of a field value.

Using Eq.~\eqref{eq_master_fieldvalue}, we estimate the power spectrum of an observed field value, which reproduces the known results on the axion field value~\cite{Foster:2017hbq}.
We can choose at $\vec x = 0$ without loss of generality since a shift of the coordinate just results in a constant phase of each partial wave.
Since $\tilde \Phi$ becomes independent of a direction of velocity in this case, we sum the index $l$ to marginalize the direction dependence.
After the summation over $l$, the field value at $\vec x=0$ is written as
\begin{align}
    \tilde\Phi(f_n)
   &=
   \frac{T}{2}
   \sigma_\phi
   ~
    \sqrt{ \Delta_s(f_n)}
    ~
    \left[
   \frac{r_n}{{\sqrt{2}}}
   \exp(i\theta_n)
   \right]
   ,
    \label{eq_phi_fj_superposed_app}
\\
     \Delta_s(f_n)
    &\equiv
     \frac{\sum_l\Delta N_{n,l}}{N_\phi}
    =
    \int^{f_n+\Delta f/2}_{f_n-\Delta f/2} 
    \overline f_\mathrm{SHM}(v)
    \frac{\df v}{\df f} \df f
    \label{eq_def_Delta_N_app}
    \\
    &=
    \frac{1}{2}\bigg[
        \text{erf}\left(\frac{v-v_\odot}{v_\mathrm{vir}}\right)
        +\text{erf}\left(\frac{v+v_\odot}{v_\mathrm{vir}}\right)
\nonumber\\&\quad
    +\frac{
    v_\mathrm{vir}
    }{\sqrt{\pi } v_\odot}
     \left(
         e^{-\frac{(v +v_\odot)^2}{v_\mathrm{vir}^2}}
     -
         e^{-\frac{(v -v_\odot)^2}{v_\mathrm{vir}^2}}
     \right)
     \bigg]^{v(f_n+\Delta f/2)}_{v(f_n-\Delta f/2)}
    ,
\end{align}
where we omit $\vec x=\vec 0$ in the argument and erf(x) represents the error function.
Note that $\Delta_s(f_n)$ appears inside the square root in Eq.~\eqref{eq_phi_fj_superposed_app} since we estimate the variance of the sum over $l$ in a similar way to derivation in Eqs.~\eqref{eq_phi_sum_before}-\eqref{eq_phi_after_sum_i}.

\begin{figure}[tb]
  \centering
  \includegraphics[width=1.0\columnwidth]{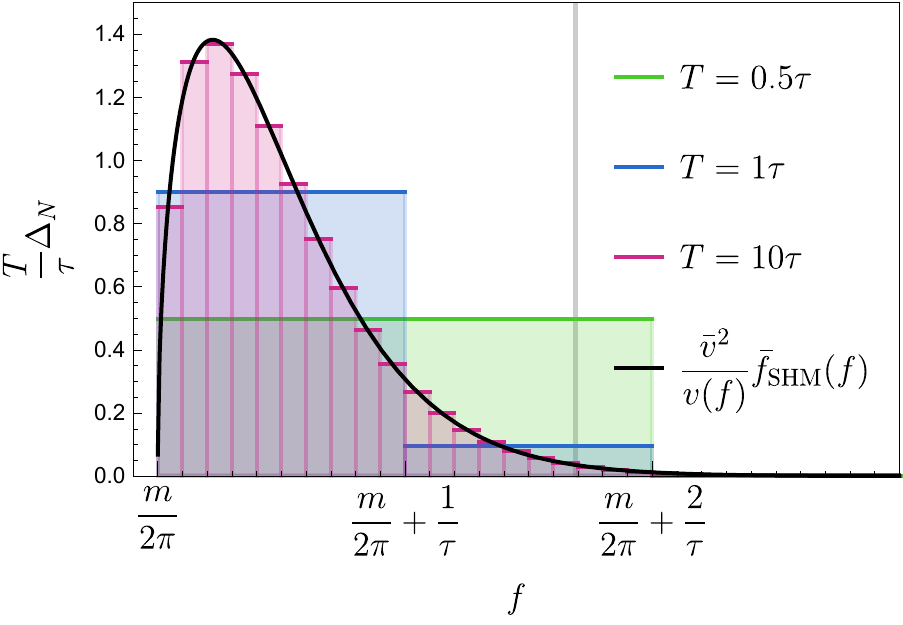}
  \caption{
  Deterministic part of the power spectrum of bosonic field amplitude $\Delta_s$ in Eq.~\eqref{eq_def_Delta_N_app}.
  A vertical axis represents the spectral shape normalized by $T/\tau$ such that the spectral shape converges to $(\bar v^2/v) \bar f_\mathrm{SHM}$ (black line) in large $T$ limit.
  The three colored lines present cases with different measurement times, $T=0.5\tau$ (green), $1\tau$ (blue) and $10\tau$ (red).
  The frequency space is discrete as $\Delta f = 1/T$ for each case.
  }
  \label{fig_powerspec_phi}
\end{figure}

We show $(T/\tau)\Delta_s(f_n)$ in Fig.~\ref{fig_powerspec_phi}, where blue, green and red lines represent the measurement time $T= 0.5\tau,~\tau,$ and $10\tau$, respectively.
For $T>\tau$, $(T/\tau)\Delta_s(f_n)$ converges to the distribution of standard halo model (black line) whose 99\% of power is contained within $f\in [m/2\pi, m/2\pi +1.69\tau^{-1}]$ (gray vertical line). 
We can approximate $\Delta_s(f_n)$ in two cases:
When the observation time is  shorter than the coherent time $T\ll \tau$, the frequency bin is so wide that one bin contains almost all DM waves, 
$\int_{\Delta f} \overline f_\mathrm{SHM}(v) \frac{\df v}{\df f} \df f \to 1$ at $f\sim m/(2\pi)$.
On the other hand, when $T\gg \tau$, the integration range is so narrow that the integration is approximated by
$\int_{\Delta f} \overline f_\mathrm{SHM}(v) \frac{\df v}{\df f} \df f
\simeq 
\overline f_\mathrm{SHM}(v)
\frac{\df v}{\df f}
\Delta f
$.
Thus, the typical value of power spectrum is given by
\begin{align}
    \Delta_s(f_n)
    \to 
    \begin{cases}
    \delta_{n,n_c}
    &\quad,\quad
    (T\ll \tau)
    \\
    \frac{1}{T}
    \overline f_\mathrm{SHM}(v_n)
    \frac{\df v}{\df f}
     &\quad,\quad
    (T\gg \tau)
    \end{cases}
    ,
    \label{eq_tilde_phi_approx}
\end{align}
where $n_c$ is an index of a bin including $f_\mathrm{DM}\equiv m/(2\pi)$, and $v_n$ is the velocity of DM at the $n$-th frequency bin. 

The spatial derivative of field value is more complicated.
Using the sum of partial waves in Eq.~\eqref{eq_master_fieldvalue}, the spatial derivative of field value is given by,
\begin{align}
    \nabla_j \tilde \Phi
    (f_n,\vec x)
    \bigg|_{\vec x=0}
   &=
   \frac{T}{2}
   \sigma_\phi
   \sum_l
   im  v_j(f_n,\vec e_l)
    \sqrt{
    \frac{\Delta N_{n,l}}{2N_\phi}
    }
    ~
    r_{n,l}
    e^{
    i\theta_{n,l}
    }.
\end{align}
Again, we perform the summation over $l$ in the same way as Eq.~\eqref{eq_phi_fj_superposed}, and we derive
\begin{align}
    \nabla_j \tilde \Phi
    (f_n,\vec x)
    \bigg|_{\vec x=0}
   &=
   \frac{T}{2}
   \sigma_\phi
   m \bar v
   ~
    \sqrt{ \Delta_{j}(f_n)  }
    ~
    \left[
    \frac{r_n}{{\sqrt{2}}}
    \exp\left(
    i\theta_{n}
    \right)
    \right]
    ,
    \\
     \Delta_{j}(f_n)
    &\equiv
    \sum_l
     \frac{ [v_j(f_n,\vec e_l)]^2}{\bar v^2}
     \frac{ \Delta N_{n,l} }{N_\phi}
\nonumber\\&
    =
    \int^{f_n+\Delta f/2}_{f_n-\Delta f/2} 
     v^2
     \frac{\df v}{\df f} \df f
    \int \df^2\Omega_e
\nonumber\\&\times
    f_{\rm SHM}(\vec v(f,\vec e) + \vec v_\odot)
    \frac{[v_j(f_n,\vec e_l)]^2 }{\bar v^2}
    ,
\end{align}
where we redefine $\theta_{n,l}$ to include an unnecessary phase, the direction integral covers whole direction, and
$\Delta_{j}(f_n)$ is the velocity-weighted number fraction of DM waves.
$\Delta_j$ depends on the direction of the solar velocity, $\vec v_\odot$.
We derive analytic formulas of $\Delta_j$ in two typical directions, the direction perpendicular and parallel  to $\vec v_\odot$, which are shown in 
Eqs.~\eqref{eq_Delta_para} and \eqref{eq_Delta_perp}, respectively.

\section{Signals of dark photon in interferometer}
\label{app_interferometer_dark_photon_signal}

Here, we derive the signals of dark photon for the interferometer search in detail.
At first, we relate the light travelling time of laser~\eqref{eq_sigmal_vector} to the displacement of mirrors~\eqref{eq_displacement_dark_photon_interaction}.
Interferometer experiments observe the difference of light traveling times between two interferometer arms defined in Eq.~\eqref{eq_sigmal_vector}~\cite{Morisaki:2020gui}
\begin{align}
    h(t) = \frac{\varphi(t,\vec e) -\varphi(t,\vec d)}{4 \pi \nu  L}.
\end{align}
The laser phase $\varphi(t,\vec e)$ depends on the round-trip time $T_r$, in which the injected light return back to the injected point.
Using the positions of an input and end-test-mass mirrors, $ x_\mathrm{in}(t)$ and $x_e(t)$, $T_r$ is written as
\begin{align}
	T_r(t)
	&=
	-x_\mathrm{in}(t) +2x_e(t-L)- x_\mathrm{in}(t-2L),
	\label{eq_round_trip_time}
\end{align}
and the laser phase is given by
\begin{align}
	\varphi(t,\vec e)
	&=
	2\pi \nu (t-T_r)
	+\varphi_0,
\end{align}
where $\varphi_0$ is a constant phase. 
The dark electric force fluctuates the position of mirrors.
Without fluctuations, the input mirror is located at $\vec x =0$ and the end-test-mass mirror is at $\vec x = L\vec e$.
With fluctuations, their positions projected on the arm direction $\vec e$ are given by
\begin{align}
	x_\mathrm{in} (t) 
	&= \vec e\cdot \delta \vec x(t,\vec x =0; (Q/M)_\mathrm{in})
	,
	\\
	x_e (t) 
	&= 
	L
	+ 
	\vec e\cdot \delta \vec x(t,L\vec e; (Q/M)_e)
	,
\end{align}
where $(Q/M)_\mathrm{in}$ and $(Q/M)_e$ are the charge mass ratios of the input and end test mass mirrors, respectively.
When ingredients are different between two mirrors, the charge-mass ratio is different, $(Q/M)_\mathrm{in}\neq (Q/M)_e$.

Next, we decompose the phase into the three different terms in the following way,
\begin{align}
	\varphi(t,\vec e) 
	&=
	\varphi_0
	+2\pi \nu(t-2L)
\nonumber\\&\quad
	-2\pi\nu(
	\delta L_\mathrm{time}
	+\delta L_\mathrm{space}
	+\delta L_\mathrm{charge}
	).
\end{align}
The first term represents the phase shift by a finite-time traveling effect, which is defined by
\begin{align}
	\delta L_\mathrm{time}
	&\equiv
	\vec e\cdot [
	-\delta \vec x(t,0; (Q/M)_\mathrm{in})
	+2\delta \vec x(t-L,0; (Q/M)_\mathrm{in})
\nonumber\\&\quad
	-\delta \vec x(t-2L,0; (Q/M)_\mathrm{in})
	]
\nonumber\\ &=
	\frac{e \epsilon_D (Q/M)_\mathrm{in}  }{m^2}
	~
	\frac{\partial }{\partial t}
	\sum_k
	e_k
	\bigg[
	- { A_k}(t,\vec 0)
	+2 { A_k}(t-L,\vec 0)
\nonumber\\&\quad
	- {A_k}(t-2L,\vec 0)
	\bigg]
\nonumber\\ &\simeq
	\frac{4e \epsilon_D (Q/M)_\mathrm{in}  }{m^2}
	\sin^2\left(\frac{mL}{2}\right)
	~
	\frac{\partial }{\partial t}
	\sum_k
	e_k
	{ A_k}(t-L,\vec 0)
	.
\end{align}
The second term is induced by the spatial difference of the DM field value, 
\begin{align}
	\delta L_\mathrm{space}
	&\equiv
	2\vec e\cdot [
	\delta \vec x(t-L,L\vec e; (Q/M)_\mathrm{in})
\nonumber\\&\quad
	-\delta \vec x(t-L,0; (Q/M)_\mathrm{in})
	]
\nonumber	\\ &\simeq
	\frac{2e \epsilon_D  (Q/M)_\mathrm{in}}{m^2} L
	~
	\frac{\partial }{\partial t}
	\sum_{k,j} e_k e_j 
	\nabla_j A_k(t-L,\vec 0)
	,
	\label{eq:deriveapp}
\end{align}
where we assume that $L|\vec k|\sim L m \bar v \ll1$, which is valid in the case of ground based interferometers.
The third term only appears when the ingredient of two mirrors are different, 
\begin{align}
	&\delta L_\mathrm{charge}
\nonumber\\&=
	2\vec e\cdot [
	\delta \vec x(t-L,L\vec e; (Q/M)_e)
\nonumber\\&\quad
	-\delta \vec x(t-L,L\vec e; (Q/M)_\mathrm{in})
	]
	\\ &\simeq
	\frac{2e \epsilon_D ((Q/M)_e-(Q/M)_\mathrm{in}) }{m^2}
	~
	\frac{\partial}{\partial t}
	\sum_k e_k
	A_k(t-L,L\vec e)
	.
\end{align}
Since each term differently depends on parameters of a detector, we consider them separately.


After the Fourier-transformation, the field values of the vector field are represented by Eqs.~\eqref{eq_tildeA} and \eqref{eq_nablatildeA}.
Then, $s_\mathrm{time}$ is given by
\begin{align}
	s_\mathrm{time}(f_n)
	&=
	\int_{-T/2}^{T/2} \df t e^{- 2\pi i f_n t}
	\frac{
		\delta L_\mathrm{time}(\vec e)
		-\delta L_\mathrm{time}(\vec d)
	}{-2L}
	\\&=
	e \epsilon_D  T
	\frac{(Q/M)_\mathrm{in}\sigma_A   }{mL}
	\sin^2\left(\frac{mL}{2}\right)
\nonumber\\&\times
    \sqrt{\Delta_s(f_n)}
	\left[
	\sum_k
	(e_k - d_k)
	~
	\frac{r_{k,n} }{{\sqrt{2}}}
	\exp(i\theta_{k,n})
	\right],
\end{align}
where we redefine $\theta_{k,n}$ to include constant phases.
Since the factor in $[...]$ is a sum of three complex Gaussian variables, we replace it by one Complex Gaussian variable with a variance $\sqrt{ \sum_k (e_k-d_k)^2} = |\vec e-\vec d|$.
Considering the above, $s_\mathrm{time}$ is written by using new stochastic variables, $r_n$ and $\theta_n$, as
\begin{align}
	s_\mathrm{time}(f_n)
	&=
	\left(
	e \epsilon_D  T
	\frac{(Q/M)_\mathrm{in}\sigma_A   }{mL}
	\sin^2\left(\frac{mL}{2}\right)
	|\vec e-\vec d|
	\right)
	~
\nonumber\\&\times
    \sqrt{\Delta_s(f_n)}
	\left[
	\frac{r_{n} }{{\sqrt{2}}}
	\exp(i\theta_{n})
	\right]
	,
\end{align}
where $\theta_{n}$ follows a uniform distribution over $[0,2\pi]$, and  $r_{n}$ follows a standard Rayleigh distribution.

In the similar way, $s_\mathrm{charge}$ is evaluated as 
\begin{align}
	s_\mathrm{charge}(f_n)
	&=
	\left(
	e \epsilon_D T
	\frac{((Q/M)_e-(Q/M)_\mathrm{in})\sigma_A }{2Lm}
	|\vec e -\vec d|
	\right) 
\nonumber\\&\times
	\sqrt{ \Delta_s(f_n)}
	\left[
	\frac{r_{n} }{{\sqrt{2}}}
	\exp(i\theta_{n})
	\right]
	.
\end{align}
Both $s_\mathrm{time}$ and $s_\mathrm{charge}$ are similar to the axion signal except for the proportional constants.


On the other hand, $s_\mathrm{space}$ depends on a frequency differently,
\begin{align}
	s_\mathrm{space}(f_n)
	&=
	e \epsilon_D  T
	\frac{ (Q/M)_\mathrm{in} \sigma_A  \bar v}{2} 
\nonumber\\&\times
	\sum_{k,j}
    (e_k e_j -d_k d_j)
	\sqrt{ \Delta_{j}(f_n)  }
	\left[
	\frac{r_{k,n} }{{\sqrt{2}}}
	\exp\left(
	i\theta_{k,n}
	\right)
	\right]
	.
\end{align}
In the following, we assume the interferometer with two orthogonal arms like LIGO, Virgo and KAGRA, and determine coordinates as ${\vec e}= (1,0,0)$ and ${\vec d}=(0,1,0)$.
In this case, $s_\mathrm{space}(f_n)$ is simplified as 
\begin{align}
	s_\mathrm{space}(f_n)
	&=
	\left( 
	e \epsilon_D  T
	\frac{ (Q/M)_\mathrm{in} \sigma_A  \bar v}{2} 
	\right) 
\nonumber\\&\times
    \sqrt{\Delta_x(f_n)+\Delta_y(f_n)}
	\left[
	\frac{r_{n} }{{\sqrt{2}}}
	\exp\left(
	i\theta_{n}
	\right)
	\right]
	.
\end{align}

For a general interferometer configuration with two arms pointed to unit vectors, $\vec e$ and $\vec d$, $s_\mathrm{space}(f_n)$  is written by
\begin{align}
	s_\mathrm{space}(f_n)
	&=
	\left( 
	e \epsilon_D  T
	\frac{ (Q/M)_\mathrm{in} \sigma_A  \bar v}{2} 
	\right) 
\nonumber\\&\times
	\sqrt{
	(\vec e\cdot \vec u)^2
	+(\vec d\cdot \vec u)^2
	-2(\vec e\cdot \vec d)(\vec e\cdot \vec u)(\vec d\cdot \vec u)
	}
\nonumber\\&\times
	\left[
	\frac{r_{n} }{{\sqrt{2}}}
	\exp\left(
	i\theta_{n}
	\right)
	\right]
	,
\end{align}
with $\vec u  \equiv \left(\sqrt{\Delta_x(f_n)}, \sqrt{\Delta_y(f_n)}, \sqrt{\Delta_z(f_n)}\right)$

\section{Likelihood in a deterministic and a stochastic case.}
\label{app_derive_likelihood}

We showed the likelihood of the total power spectrum $\rho$~\eqref{eq:freq_range} in Sec.~\ref{sec_likelihood}.
Here, we derive the likelihood for a deterministic case~\eqref{eq_likelihood_deterministic} and a stochastic case~\eqref{eq_label_likelihood}.

First, we consider the deterministic likelihood of $\rho$ neglecting the  Rayleigh distribution of the field value: 
\begin{align}
    &\mathcal{L}(\rho | \{\lambda_n\}) 
\nonumber\\ 
    &\equiv
    \int
	\left(
	\prod_n^{N_\mathrm{bin}}
	{\rm d} \rho_n~
	{\mathcal L} (\rho_n|\sqrt{2}\lambda_n)
	\right)
	\delta\left(\rho - \sum_n^{N_\mathrm{bin}} \rho_n  \right)
\nonumber\\&=
     \int 
    \left(
    \prod_{n}^{{N_\mathrm{bin}}} 
    \frac{\df^2 \hat{\mathcal{N}}_n}{2\pi} ~e^{-|\hat{\mathcal{N}}_n|^2/2}
    \right)
    \int^{\infty}_{-\infty} \frac{\df k}{2\pi}
\nonumber\\&\times
    \exp\left[
        ik\left(
            \rho - 
            \sum_n^{N_\mathrm{bin}}\left| \hat{\mathcal{N}}_n + \sqrt{2}\lambda_n ~e^{i\theta_n} \right|^2 
        \right)
    \right]
\nonumber\\&=
    e^{-\Lambda/2}
    \int \frac{\df k}{2\pi}
    e^{ik\rho}
    \sum_m
    \frac{1}{m!}
    \frac{1}{(1+2ik)^{{N_\mathrm{bin}}+m} }
    \left(\frac{\Lambda}{2}\right)^m
\nonumber\\&=
    \frac{e^{-(\rho+\Lambda)/2}}{2}
    \sum_m
    \frac{1}{m!(m+{N_\mathrm{bin}}-1)!}
    \left(\frac{\Lambda}{2}\right)^m
    \left(\frac{\rho}{2}\right)^{{N_\mathrm{bin}}+m-1}
\nonumber\\&=
    \frac{e^{-(\rho+\Lambda)/2}}{2}
    \left(\frac{\rho}{\Lambda }\right)^{\frac{{N_\mathrm{bin}}-1}{2}}
    I_{{N_\mathrm{bin}}-1} (\sqrt{\Lambda\rho}),
\end{align}
where we define the sum of signals
\begin{align}
    \Lambda \equiv 2\sum_n^{N_\mathrm{bin}}(\lambda_n)^2,
\end{align}
and we use the modified Bessel function of the first kind  $I_{n-1}$, which satisfies 
\begin{align}
    \sum_{m=0}^\infty 
    \frac{1}{m!(m+n-1)!}
    x^m
    =
    x^{\frac{1-n}{2}} I_{n-1}( 2\sqrt{x}).
\end{align}
This is the noncentral chi-square distribution when the number of degrees of freedom {is} $2{N_\mathrm{bin}}$, and the noncentrality parameter $\Lambda$.

Next, we estimate the stochastic likelihood marginalising the amplitude of the field value:
 \begin{align}
	\overline{\mathcal L}(\rho|\{\lambda_n\})
	&=
	\int
	\left(
	\prod_{l }^{N_\mathrm{bin}}
	{\rm d} \rho_n~
	\overline{\mathcal L} (\rho_n|\lambda_n)
	\right)
	\delta\left(\rho - \sum_n^{N_\mathrm{bin}} \rho_n  \right)
 \nonumber\\& =
    \sum_n^{N_\mathrm{bin}}  
     \frac{
    w_n
     }{{2}(1+\lambda_n^2)}
      \exp\left( 
     - \frac{\rho}{{2}(1+\lambda_n^2)}
     \right)
	,
	\\
    w_n
	&\equiv
	\prod_{n'(\neq n)}^{N_\mathrm{bin}}
	\frac{1+\lambda_n^2}{\lambda_n^2-\lambda_{n'}^2},
\end{align}
where we assume $\lambda_n\neq \lambda_n'$ for all $n\neq n'$.
To evaluate the integration, we transform the delta function as $\delta(x)=\int\df k/(2\pi) e^{ikx}$, perform the integration on $\rho_n$, and then evaluate the integration on $k$ by the saddle point method.
Assuming each $\lambda_n$ has a different value, $N_\mathrm{bin}$ distinct saddle points appear.
On the other hand, when all {$\lambda_n$ have the same value $\lambda_n= \lambda$, }
these saddle points degenerate and the likelihood is {computed by the $N_\mathrm{bin}$-th residue as},
\begin{align}
	&\overline{\mathcal L}(\rho|\{\lambda_n=\lambda\})
\nonumber\\
&\sim 
    \frac{1}{(N_\mathrm{bin}-1)!}
    \frac{\rho^{N_\mathrm{bin}-1}}{
        {2^{N_\mathrm{bin}}}
        (1+\lambda^2)^{N_\mathrm{bin}}
    }
     \exp\left( 
     -\frac{\rho}{{2}(1+\lambda^2)}
     \right)
     \label{eq_likelihood_same_sigma}
   .
\end{align}

\section{Approximate formula for short- and long-time measurements}
\label{sec_approx_measurmetn}

We derive the analytic formula for the upper bound $\bar \lambda_\mathrm{up}$~\eqref{eq_def_lambdaup} in two limiting cases, $N_\mathrm{bin}=1$ and $N_\mathrm{bin}\gg 1$.
We use the detection threshold $\rho_\mathrm{dt}$~\eqref{eq_def_rhodt} instead of the observed data $\rho_\mathrm{obs}$, where one can easily perform the similar analysis on $\rho_\mathrm{obs}$.
Here, we choose the analyzed frequency range $\kappa$ in Eq.~\eqref{eq:freq_range} large enough to cover almost all DM distributions for simplicity.
We discuss the optimal choice of $\kappa$ later.

When the measurement time is much smaller than the coherent time $T\ll \tau$, the DM signal only appears at a single frequency bin, that is, $N_\mathrm{bin}=1$.
Then, the spectral shape of the DM signal is given by
\begin{align}
    \Delta_X(f_n)
    \to
    \Delta_{X,\mathrm{tot}} \delta_{n,n_c}
    \quad\mathrm{for}\quad
    T<\tau/\kappa,
\end{align}
where $n_c$ is the index of the bin that includes $f_\mathrm{DM}$, and $T<\tau/\kappa$ is the condition for $N_\mathrm{bin}=1$ found in Eq.~\eqref{eq_num_bin}.
{$\Delta_{X,\mathrm{tot}}$ is the sum of all $\Delta_X (f_n)$ over the whole frequency range $n$ as defined in Eqs.~\eqref{eq_delta_N_tot} and \eqref{eq_delta_pp_tot},
and their values are given by $\Delta_{s,\mathrm{tot}} = 1$, $\Delta_{\perp,\mathrm{tot}} \simeq 0.19, \Delta_{\parallel,\mathrm{tot}} \simeq 0.62$.}
Note that {$\Delta(f_n)$ becomes proportional to Kronecker delta $\delta_{n,n_c}$,}
since the frequency resolution is so rough that the single frequency bin covers almost all DM particles.
In this $N_\mathrm{bin}=1$ case, the likelihood function is given by Eq.~\eqref{eq_barL_single}.
The detection threshold defined in Eq.~\eqref{eq_def_rhodt} is calculated as~\cite{Centers:2019dyn}
\begin{align}
	 \rho_\mathrm{dt}
	 = {2}\ln(\alpha^{-1}).
\end{align}
The upper bound on $\bar\lambda$ is given by Eq.~\eqref{eq_def_lambdaup} as
\begin{align}
    1-\beta 
    &= 
	1-
	\exp\left(
	-\frac{\rho_\mathrm{dt}}{
	    {2}+{2}(\bar\lambda_\mathrm{up})^2  
	    \Delta_{X,\mathrm{tot}}
	    }
	\right),
	\\
	\bar\lambda_\mathrm{up}
	&=
	\frac{1}{\sqrt{ \Delta_{X,\mathrm{tot}}}}
	\sqrt{\frac{\rho_\mathrm{dt}}{-{2}\ln \beta} -1}
	\simeq 
	\frac{1}{\sqrt{ \Delta_{X,\mathrm{tot}}}}
	\sqrt{\frac{\ln(\alpha)}{\ln(\beta)}-1}.
\end{align}
Using Eq.~\eqref{eq_def_barlambda}, we relate $\bar\lambda_\mathrm{up}$ to the upper bound on the coupling constant as
\begin{align}
    \epsilon_{D,\mathrm{up}}, g_{a,\mathrm{up}}(T)
    &\propto 
    \sqrt{\frac{\ln(\alpha)}{\ln(\beta)}-1} 
     \frac{ \sqrt{T S_{\rm noise}(f_\mathrm{DM})}}{
     2  \sqrt{ \Delta_{X,\mathrm{tot}} } T 
     }
\nonumber\\&     \propto
     T^{-1/2}
    \quad\mathrm{for}\quad 
    T<\tau/\kappa.
    \label{eq_gth_short}
\end{align}
 This formula explicitly shows the improvement of the upper bound by time for a short measurement time.

On the other hand, when the measurement time is much larger than the coherent time $T\gg \tau$, the DM signal  spreads over $N_\mathrm{bin}(\gg 1)$ frequency bins. 
In this case, we can approximate the probability distribution of $\rho$ by a Gaussian distribution in the following way.
The likelihood in each frequency bin is given by Eq.~\eqref{eq_barL_single}. 
Let us consider a group of bins whose frequencies lie within a certain range, $f<f_n<f+\delta f$.
We choose a sufficiently small $\delta f$ so that the differences in the likelihood function between the bins in the group are negligible. In the large $N_\mathrm{bin}$ limit, the frequency interval $\Delta f=1/T$ becomes infinitesimally small and the number of bins in the group blows up.
Then, by virtue of the central limit theorem, the sum of $\rho_n$ in the bin group follows Gaussian distribution.
Based on the likelihood~\eqref{eq_barL_single}, 
the mean value and variance of the summed $\rho_n$ are given by $\mathrm{E}[\rho_n]= {2}(1+\lambda_n)^2$ and $\mathrm{Var}[\rho_n] = {4}(1+\lambda_n^2)^2$, respectively.
The total sum, $\rho =\sum_n^{N_\mathrm{bin}} \rho_n$, also follows the Gaussian distribution with a mean value $ \mu_\rho\equiv\mathrm{E}[\rho]= \sum_n^{N_\mathrm{bin}} \mathrm{E}[\rho_n]$ and a variance $\sigma^2_\rho\equiv \mathrm{Var}[\rho]= \sum_n^{N_\mathrm{bin}} \mathrm{Var}[\rho_n]$.
The likelihood of $\rho$ is approximated by
\begin{align}
	\overline{\mathcal L}(\rho|\{\lambda_n\})
	&\to 
	\frac{1}{\sqrt{2\pi \sigma^2_\rho } }
	\exp\left(
	-
	\frac{(\rho - \mu_\rho)^2}{\sigma^2_\rho}
	\right)
	\quad\mathrm{for}\quad
    T\gg\tau,
    \label{eq_approx_likelihood}
    \\
    \mu_\rho(\bar \lambda_X)
    &=
    \sum_n  (1+\lambda_n^2) 
    = 
    {2} ( 
    N_\mathrm{bin}
    +\bar \lambda_X^2 \Delta_{X,\mathrm{tot}}
    )
    ,
    \\
     \sigma^2_\rho(\bar \lambda_X)
    &=
    {4}\sum_n (1+\lambda_n^2)^2
\nonumber\\&    \simeq 
    {4}(
    N_\mathrm{bin}
    +2\bar \lambda_X^2 \Delta_{X,\mathrm{tot}}
     +\sum_n \lambda_n^4
    ) 
    ,
\end{align}
where we use  $\lambda_n \equiv  \bar \lambda_X \sqrt{\Delta_X(f_n)}$ and $\sum_n^{N_\mathrm{bin}} \Delta_X(f_n) = \Delta_{X,\mathrm{tot}}$.
Now we can compute the upper bound by plugging the approximated likelihood function~\eqref{eq_approx_likelihood} into Eq.~\eqref{eq_def_lambdaup}.

For later convenience, it is useful to define {a function $M_\chi$ which denotes} the relation between the peak width and the area of the Gaussian distribution as
\begin{align}
    \chi  = 
    \int_{M_\chi}^\infty \df z 
    \frac{1}{\sqrt{2\pi}} \exp(-z^2/2),
\end{align}
where $\chi$ is a probability where a standard Gaussian variable exceeds $M_\chi(>0)$. 
{Note that $M_\chi$ is the inverse function of the complementary error function.}
For example, $M_{0.32} \simeq 0.468$, $M_{0.05} \simeq 1.64$, and $M_{0.01} \simeq 2.33$.
Given that the likelihood in Eq.~\eqref{eq_def_rhodt} is well approximated by a Gaussian function~\eqref{eq_approx_likelihood}, 
the threshold $\rho_\mathrm{dt}$ reads
\begin{align}
    \rho_\mathrm{dt}
    &=
    \mu_\rho(\bar \lambda_X=0) + M_\alpha \sigma_\rho(\bar \lambda_X=0) 
    =
    {2}
    (N_\mathrm{bin}
    +M_\alpha\sqrt{N_\mathrm{bin}}
    )
    .
\end{align}
In the same way, Eq.~\eqref{eq_def_lambdaup} is evaluated as
\begin{align}
    \rho_\mathrm{dt} 
    &=
    \mu_\rho(\bar\lambda_\mathrm{up} ) - M_{1-\beta} \sigma_\rho(\bar\lambda_\mathrm{up} )
\\&\simeq 
    2 ( 
    N_\mathrm{bin}
    +\bar \lambda_\mathrm{up}^2 \Delta_{X,\mathrm{tot}}
    )
    - 4M_{1-\beta}
    (
     N_\mathrm{bin}
    +2\bar \lambda_\mathrm{up}^2 \Delta_{X,\mathrm{tot}}
    )
    \label{eq_approx_lambdaup_euality}
    .
\end{align}
In the last line, we neglect $\sum_n \lambda_n^4$ term since it is subdominant in a large $N_\mathrm{bin}$ limit as we will confirm later.

\begin{figure}[t]
  \centering
  \includegraphics[width=1.0\columnwidth]{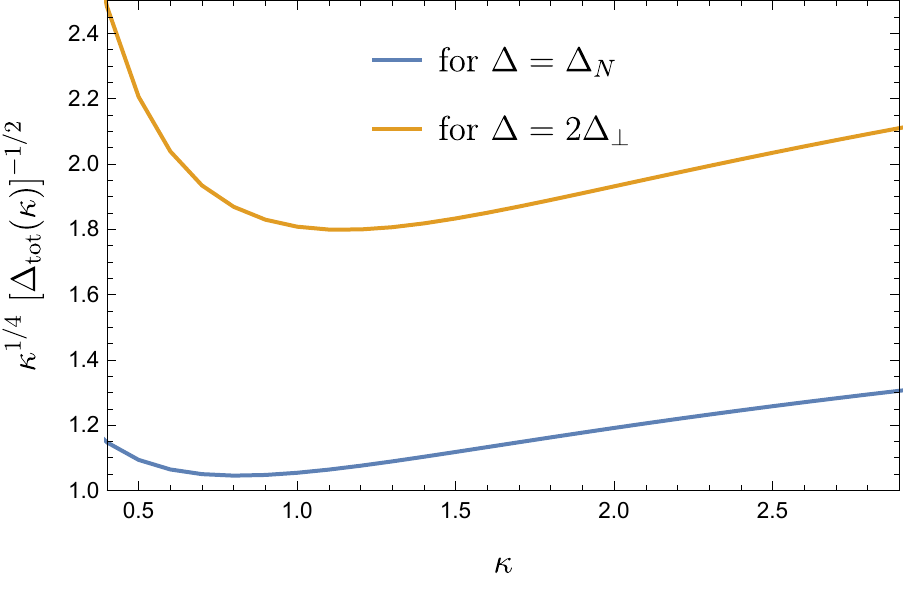}
  \caption{
  $\kappa$ dependence of the upper bound on the coupling constant for $N_\mathrm{bin}\simeq \kappa T/\tau\gg 1$, $\epsilon_{D,\mathrm{up} }, g_{a,\mathrm{up}}\propto \bar\lambda_\mathrm{up} \propto (\Delta_{X,\mathrm{tot}})^{-1/2} \kappa^{1/4}$ (see Eqs.~\eqref{eq_def_barlambda} and \eqref{eq_lambdas_twolimit}).
   $\kappa$ determines the frequency range $[f_\mathrm{DM}, f_\mathrm{DM}+\kappa/\tau]$ of our analysis defined in Eq.~\eqref{eq:freq_range}.  
  The blue and orange lines represent  the cases with the velocity-independent signal $\Delta=\Delta_s$ and the velocity-dependent signal of the conservative direction $\Delta_X= 2\Delta_\perp$ which has been discussed below Eq.~\eqref{eq_sspace}, respectively.
  For $\kappa\lesssim 1$,  $\Delta_{X,\mathrm{tot}}$ increases and results in a tighter constraint as $\kappa$ increases. However, for a too large $\kappa \gtrsim 2$, the analysis includes an additional frequency range where DM signal is very little and ends up weakening the constraint. 
  }
  \label{fig_kappa_Optimize}
\end{figure}

Combining both relations, $\bar\lambda_\mathrm{up} $ is estimated as
\begin{align}
    \bar\lambda_\mathrm{up} ^2
    \simeq \frac{1}{\Delta_{X,\mathrm{tot}}}
    (M_\alpha+ M_{1-\beta})\sqrt{N_\mathrm{bin}}
    .\label{eq:lambdaup_N>>1}
\end{align}
Here, we revisit the validity of approximation in Eq.~\eqref{eq_approx_lambdaup_euality}.
In a large $N_\mathrm{bin}$ limit, $\sum_n \bar \lambda_n^4$ is subdominant because of 
$\bar \lambda_\mathrm{up}\propto N_\mathrm{bin}^{1/4}$, 
$\Delta_X(f_n)\propto 1/T \propto N_\mathrm{bin}^{-1}$, 
and $\sum_n \bar \lambda_n^4 = \sum_n \bar \lambda_\mathrm{up}^4 [\Delta_X(f_n)]^2 \propto N^{0}$.
Finally, the upper bound of the coupling constant is given by
\begin{align}
    \epsilon_{D,\mathrm{up}}, g_{a,\mathrm{up} }(T)
    &\propto 
     \sqrt{\frac{
        (M_\alpha+ M_{1-\beta})
        \sqrt{N_\mathrm{bin}}
     }{\Delta_{X,\mathrm{tot}} }}
     \frac{ \sqrt{T S_{\rm noise}(f_\mathrm{DM})}}{2  T \gamma }
\nonumber\\& \propto
     (\kappa T/\tau)^{1/4} T^{-1/2}
     \propto
     T^{-1/4}
    \quad\mathrm{for}\quad T\gg\tau.
    \label{eq_gth_long}
\end{align}
Here we explicitly showed that the slow-down of improvement by time holds even in the case with DM fluctuations.
In summary, we derived two analytical formulas of $\bar\lambda_\mathrm{up} (T)$ for $\tau<T$~\eqref{eq_gth_short} and $T\gg \tau$~\eqref{eq_gth_long} as
\begin{align}
    \bar\lambda_\mathrm{up} (T)
    &=
    \begin{cases}
        (\Delta_{X,\mathrm{tot}})^{-1/2}
          \sqrt{\frac{\ln(\alpha)}{\ln(\beta)}-1} 
        \\\qquad\mathrm{for}\quad T<\tau/\kappa
        \\
        (\Delta_{X,\mathrm{tot}})^{-1/2}
        \sqrt{M_\alpha+ M_{1-\beta}} \left(\kappa T/\tau \right)^{1/4}
        \\
        \qquad\mathrm{for}\quad T\gg\tau
    \end{cases}
    .
    \label{eq_analytic_limit_lambdaup_app}
\end{align}

In the above discussions, we limit the frequency range of data as $f_\mathrm{DM}< f_n <f_\mathrm{DM}(1+\kappa \bar v^2)$. 
Here, we investigate the optimized choice of $\kappa$ and its effect on the sensitivity by using the analytic formula \eqref{eq_lambdas_twolimit}.
We consider the long measurement case $T\gg \tau$, in which the number of bins is given by  $N_\mathrm{bin} \simeq \kappa T/\tau$.
The upper limit of the coupling constant depends on $\kappa$ through 
$\bar\lambda_\mathrm{up} \propto (\Delta_{X,\mathrm{tot}})^{-1/2} \kappa^{1/4}$ 
as shown in Eq.~\eqref{eq_analytic_limit_lambdaup_app}.
When we use a small $\kappa$, $\Delta_{X,\mathrm{tot}}$ depends on $\kappa$ through the integration range of a frequency in the following way:
\begin{align}
     \Delta_{s,\mathrm{tot}}(\kappa)
    &=
    \int^{f_\mathrm{DM}+\kappa/\tau}_{f_\mathrm{DM}} 
    \overline f_\mathrm{SHM}(v)
    \frac{\df v}{\df f} \df f
    ,
\end{align}
for the velocity-independent signal.
When $\kappa$ is much smaller than unity, the measurement covers an only small fraction of the DM signal, which leads to smaller $\Delta_{X,\mathrm{tot}}$.
Since $g_\mathrm{up}\propto \bar\lambda_\mathrm{up} \propto (\Delta_{X,\mathrm{tot}})^{-1/2} \kappa^{1/4}$,
{we present $(\Delta_{X,\mathrm{tot}})^{-1/2} \kappa^{1/4}$} in Fig.~\ref{fig_kappa_Optimize} for a velocity-independent signal (blue line) and a velocity-dependent signal (orange line). 
Although the upper bound becomes the smallest for $\kappa \simeq 1$, the choice of $\kappa$ does not dramatically change the upper bound and less important compared to the amplitude fluctuation on $T\gtrsim \tau$.
In this paper,  in order to gather up most of the signal, we adopt a value of $\kappa$ which covers 99\% of the dark matter signal: $\kappa \simeq 1.69$ for the velocity-independent signals and $\kappa \simeq 2$ for the velocity-dependent signals.


\bibliographystyle{apsrev4-2}
\bibliography{Ref}

\end{document}